\pdfoutput=1
\documentclass[aip,pof,amsmath,amssymb,
reprint
]{revtex4-1}

\usepackage{dcolumn}
\usepackage[colorlinks,
            linkcolor=blue,
            urlcolor=blue,
            anchorcolor=blue,
            citecolor=blue
            ]{hyperref}

\usepackage{bm}
\usepackage{graphicx}
\graphicspath{{./images/}{./bubble/}{./bubbleMach/}{./cylinders/}{./results/}}
\usepackage{epstopdf}

\begin{document}


\title{A non-oscillatory energy-splitting method for the computation of compressible  multi-fluid flows}

\author{Xin Lei}
 \affiliation{School of Mathematical Sciences, Beijing Normal University, Beijing 100875, People's Republic of China}
 \email{xinlei@mail.bnu.edu.cn}
\author{Jiequan Li}
 \affiliation{Laboratory of Computational Physics, Institute of Applied Physics and Computational Mathematics, Beijing 100088,  People's Republic of China}
 \email{li\_jiequan@iapcm.ac.cn}
\date{\today}

\begin{abstract}
This paper proposes a new non-oscillatory {\em energy-splitting} conservative algorithm for computing multi-fluid flows in the Eulerian framework.  In comparison with   existing multi-fluid algorithms in literatures, it is shown that the mass fraction model with isobaric hypothesis is a  plausible choice  for designing numerical methods for multi-fluid flows.  Then we construct a conservative Godunov-based scheme with the high order accurate extension by using the generalized Riemann problem (GRP) solver,  through  the detailed analysis of kinetic energy exchange when fluids are mixed under the hypothesis of  isobaric equilibrium.  Numerical experiments are carried out for the shock-interface interaction and shock-bubble interaction problems, which  display the excellent performance of this type of schemes and demonstrate that  nonphysical  oscillations are suppressed  around material interfaces substantially. 

\end{abstract}

\maketitle


\section{Introduction}

It is noticed that in the computation of compressible multi-fluid flows, there are usual difficulties due to nonphysical oscillations generated at material interfaces when conservative schemes are used. This phenomenon can be unfolded by a simple example on a Cartesian structural mesh using a standard shock-capturing method. A moving material interface is initially aligned on the right boundaries of a row of cells, and separates the computational domain into two parts with different fluid materials, but with uniform pressure and non-zero velocity. Due to different thermal equations of state (EOS) for the two fluids, numerical errors of pressure and velocity may be produced and propagate away from the material interface between the forward and backward characteristic waves. Fixing the present Cartesian coordinates and fluid data, and rotating the structural mesh, it is found that this oscillatory phenomenon never disappears. In addition, if there is a shock wave interacting with the interface, density oscillations may be enlarged. Due to the possible presence of shocks in compressible multi-fluid flows, conservative schemes are appealing in the practical applications.

A slew of results on this subject were listed in Abgrall and Karni's review article\cite{abgrall_computations_2001}.
There were two typical  frameworks for numerically simulating multi-fluid dynamics: the front-tracking method and front-capturing method. The front-tracking method takes discontinuities (including material interfaces) as moving fronts, which preserve the sharpness of interfaces. The solution of the associated Riemann problem across fronts gives an indication of the motion of the fronts\cite{chern_front_1986,grove_anomalous_1990,leveque_two-dimensional_1996}.
The front tracking method eliminates the numerical diffusion and reduces post-shock oscillations common to shock-capturing methods\cite{holmes_numerical_1995}.
 An example is that Cocchi and Saurel proposed a front-tracking method  consisting of a predictor step and a corrector step in order to prevent spurious oscillations near interfaces\cite{cocchi_riemann_1997}.

The front-capturing method (shock-capturing method in this paper) simulates multi-fluids using the integral  (finite volume) form  of the governing equations. In order to identify each fluid, we need to couple the Euler equations with the $\gamma$ (ratio of specific heats)-model\cite{abgrall_how_1994,karni_multicomponent_1994,shyue_efficient_1998}, the volume fraction model\cite{shyue_efficient_1998}, the mass fraction model\cite{banks_high-resolution_2007,larrouturou_how_1991,quirk_dynamics_1996}  or the level-set model\cite{karni_multicomponent_1994,karni_hybrid_1996,mulder_computing_1992}.
For these models, some single-fluid algorithms and quasi-conservative or non-conservative approaches were suggested to ensure the correct numerical fluid mixing rules at interfaces.
For example, a non-conservative $\gamma$-model, using small viscous correction terms to remove leading-order conservation errors, was presented in terms of  primitive variables\cite{karni_multicomponent_1994}.  Then some quasi-conservative approaches was designed for  the energy equation and the $\gamma$-model\cite{abgrall_how_1994,shyue_efficient_1998},  in order to deal with strong shocks and prevent pressure oscillations through the interfaces. For the volume or mass fraction models, it is natural to ask how these models are closed or  how to compute $\gamma$ through the mass fractions or the volume fractions. In Ref.~\onlinecite{allaire_five-equation_2002}, two different closure laws were proposed with detailed mathematical analysis of the properties of the resulting models: isobaric and isothermal closures. Applying the isobaric and isothermal closures, Banks et al. introduced a high-resolution Godunov-type method with a total energy correction which is based on a uniform-pressure-velocity (UPV) flow. Another simple correction of the internal energy inside computational cells was proposed in Ref.~\onlinecite{jenny_correction_1997} to avoid spurious pressure oscillations near material interfaces. These two methods based on energy corrections do not conserve the total energy generally, unless  fluids are in thermal equilibrium. The level-set method was extensively adopted in the simulation of multi-fluid flows, using the sign of a function to identify different fluids. This  method, combined with  some   nonconservative techniques to reduce non-physical  pressure oscillations around interfaces\cite{karni_hybrid_1996,karni_multicomponent_1994}.  The ghost fluid method (GFM)\cite{fedkiw_non-oscillatory_1999} was a representative of the  modified level set method with excellent numerical  performance and an adaptive mesh refinement extension of GFM can be found in Ref.~\onlinecite{nourgaliev_adaptive_2006}.
GFM can also  be used to define interface conditions in the front-tracking method\cite{terashima_front-tracking/ghost-fluid_2009}.

There are many other multi-fluid approaches available in literature, such as the volume-of-fluid (VOF) method \cite{colella_multifluid_1989,miller_high-order_1996,ton_improved_1996}, the moment-of-fluid (MOF) method\cite{dyadechko_moment--fluid_2005} and the BGK-based model\cite{xu_bgk-based_1997}. The VOF method solves the evolution equations for volume fractions, the mass and energy equations for individual fluids, and the momentum equation for the fluid mixture, and it was improved in Ref.~\onlinecite{ton_improved_1996} to eliminate pressure oscillations by defining mixtures inside each cell with different temperatures and solving an extra  one-phase energy equation.  Ton's approach relieved the need to solve the evolution equations for volume fractions.  The MOF method, another numerical approach,  was proposed\cite{ahn_moment--fluid_2009} by using the volume fraction and centroid for a more accurate representation of the material configuration, interfaces and volume advection. 

Lagrangian and ALE frameworks are preferable in the compressible multi-fluid flows due to their sharp capturing ability of material interfaces. There are a lot of contributions on Lagrangian and ALE schemes,  see Refs.~\onlinecite{kamm_pressure_2009,hirt_arbitrary_1974,galera_two-dimensional_2010} and references therein. Still, an indispensable ingredient has to be added in order to avoid the occurrence of pressure oscillations  at material interfaces, such as  a pressure relaxation technique in the Lagrangian Godunov method with Tipton's closure model\cite{kamm_pressure_2009}. A comparative study of multi-fluid Lagrangian and Eulerian methods was made in Ref.~\onlinecite{francois_comparative_2013}, and a relaxation-projection method by Lagrangian plus remapping the flow quantities to the Eulerian  mesh was designed in Ref.~\onlinecite{saurel_relaxation-projection_2007}.

It is worth noting that any non-conservative scheme may converge to wrong solutions\cite{hou_why_1994},  providing incorrect internal energy or shock wave positions. Hence, researchers hope to simulate compressible  multi-fluid flows using conservative schemes in the Eulerian framework. For this purpose, we propose an algorithm in this paper, based on the Godunov method, that can prevent non-physical oscillations at interfaces without sacrificing the numerical results of the compressible flow phenomena involving shock waves and rarefaction waves. 
Inspired by Ref.~\onlinecite{ton_improved_1996}, we solve an equation of mass fraction and a one-phase energy equation coupling with the Euler equations. In order to simulate kinetic energy more accurately, we add a one-phase momentum equation for correction. As far as the interaction of shocks and interfaces is concerned, the exchange of kinetic  energy is processed. Motivated by Ref.~\onlinecite{jenny_correction_1997}, we use the isobaric equilibrium condition to compute the volume fractions and the ratios of specific heats inside mixed fluid cells.
A benefit of this method is its conservative form, which allows to suit for the finite volume framework.  The resulting scheme is of  Godunov-type,   and an second order accurate extension is made by using the space-time coupled generalized Riemann problem (GRP) solver \cite{ben-artzi_direct_2006,ben-artzi_generalized_2003}.  The reason of making  a choice is  the inclusion of thermodynamics into the scheme\cite{li_thermodynamical_2017}.   It is theoretically shown that the non-oscillatory property in pressure can be preserved across material interfaces. Although this paper takes two-fluid flows to illustrate the method, the proposed  algorithms can be applied to  multi-fluid flow models even when computed over unstructured meshes.

To demonstrate the performance of the proposed schemes, we carry out several numerical experiments typically for illustrating the simulation of compressible fluid flows.  They are a two-fluid compression problem in order to show the necessity of kinetic energy exchange during fluid mixture; the shock-interface interaction, and the shock-bubble interaction problems. 

We organize this paper as follows. In Section \ref{Sec:Phy-model}, we  describe basic models for immiscible compressible multi-fluid flows and discuss the cause of  pressure errors. We provide the motivation of our non-oscillatory scheme in Section \ref{Sec:En-split}, and propose the  numerical method in Section \ref{Sec:Non-Osc}.  To display the performance of the current method, we provide several typical numerical results in the context of  multi-fluid flows  including the interaction of shock-interface and the interaction of shock-bubble in Section \ref{Sec:num_res}. 

\section{\label{Sec:Phy-model}Physical models for two-fluid flows}

Under the assumption that all fluid variables are described by a single density $\rho$, a single  pressure $p$ and a common fluid velocity $\bm{u}$, the Euler equations representing conservation of mass, momentum and energy for inviscid compressible multi-fluid flows take the form 
\begin{equation}\label{density_eq}
\frac{\partial }{\partial t}\rho + \nabla \cdot (\rho \bm{u}) = 0,
\end{equation}
\begin{equation}\label{momentum_eq}
\frac{\partial}{\partial t}(\rho \bm{u}) + \nabla \cdot (\rho \bm{u}\otimes \bm{u} + p \bm{I}) = 0,
\end{equation}
\begin{equation}\label{energy_eq}
\frac{\partial}{\partial t}\left[\rho \left(e+\frac{1}{2} |\bm{u}|^2\right)\right] + \nabla \cdot \left[\rho \bm{u}\left(e+\frac{1}{2} |\bm{u}|^2\right)+p\bm{u} \right] = 0,
\end{equation}
where $e$ is the specific internal energy specified by an EOS for the mixture. 
For two immiscible fluids $a$ and $b$, we denote $z_k, \phi_k, \rho_k, p_k, e_k,  T_k, C_{v,k}, \gamma_k$ as their volume fraction, mass fraction, density, pressure, specific internal energy, temperature, specific heat capacity at constant volume and ratio of specific heats, respectively with $k=a,b$. Principles of volume average  tell us of the mixture rule of thermodynamical parameters
\begin{equation}\label{overall_rhoe}
\rho = z_a \rho_a + z_b \rho_b,\quad \rho e = z_a \rho_a e_a +z_b \rho_b e_b.
\end{equation}
In light of  the fact
\begin{equation}\label{fact}
z_a+z_b =1,\quad \rho \phi_i = \rho_i z_i,
\end{equation}
the mixture rule is written as  
\begin{equation}\label{phi_e}
\phi_a + \phi_b =1,\quad \phi_a e_a + \phi_b e_b = e.
\end{equation}
In this paper, we assume that fluids are modeled by the EOS for ideal gases
\begin{equation}
p_k = (\gamma_k -1)\rho_k e_k.
\end{equation}
Under Dalton's law of partial pressures
\begin{equation}
p = z_a p_a + z_b p_b,
\end{equation}
the mixture EOS has the unified form
\begin{equation}\label{ideal_EOS}
p = (\gamma -1)\rho e,
\end{equation}
where $\gamma$ is the effective ratio of specific heats for the  mixture given by
\begin{equation}\label{gamma}
\gamma = \gamma(\phi_a, e_a, e_b) = \frac{\phi_a e_a \gamma_a + \phi_b e_b \gamma_b}{\phi_a e_a + \phi_b e_b}.
\end{equation}
So far, various methods computing $\gamma$ had been proposed in literatures, e.g. Refs.~\onlinecite{abgrall_computations_2001,ton_improved_1996}, depending on different model assumptions. In the following, we sketch  two typical cases. 

\subsection{\label{sec:isothermal}Mass fraction model with isothermal hypothesis}
For the ideal gas $k$, $C_{v,k}$ is assumed to  depend only on  temperature, and the thermal EOS is
\begin{equation}
e_k = C_{v,k} T_k.
\end{equation}
Following  Refs.~\onlinecite{jenny_correction_1997,banks_high-resolution_2007,larrouturou_how_1991,quirk_dynamics_1996}, an isothermal hypothesis $T_a = T_b$ is used to express $\gamma$ in Eq.~\eqref{gamma} explicitly by 
\begin{equation}\label{gamma_mix}
\gamma = \gamma(\phi_a) = \frac{\phi_a C_{v,a} \gamma_a + \phi_b C_{v,b} \gamma_b}{\phi_a C_{v,a} + \phi_b C_{v,b}}.
\end{equation}
The equation of mass conservation for fluid $a$ is
\begin{equation}\label{MI_eq}
\frac{\partial}{\partial t}(\rho \phi_a) + \nabla \cdot (\rho \phi_a \bm{u}) = 0.
\end{equation}
Therefore, $\gamma$ is representable for the mixture through the solution of this one-phase mass conservation equation.  This, together with the Euler equations \eqref{density_eq}-\eqref{energy_eq}, gives a four-equation model in conservation form. It is typical to use a conservative shock-capturing scheme, such as the Godunov scheme, to numerically solve this model. Let us simulate a material interface separating two fluids with different temperatures and ratios of specific heats.   In the Eulerian framework, a moving material interface enters the interior of some computational cells. 
\begin{figure}[ht]
\centering
\includegraphics[width=0.4\linewidth]{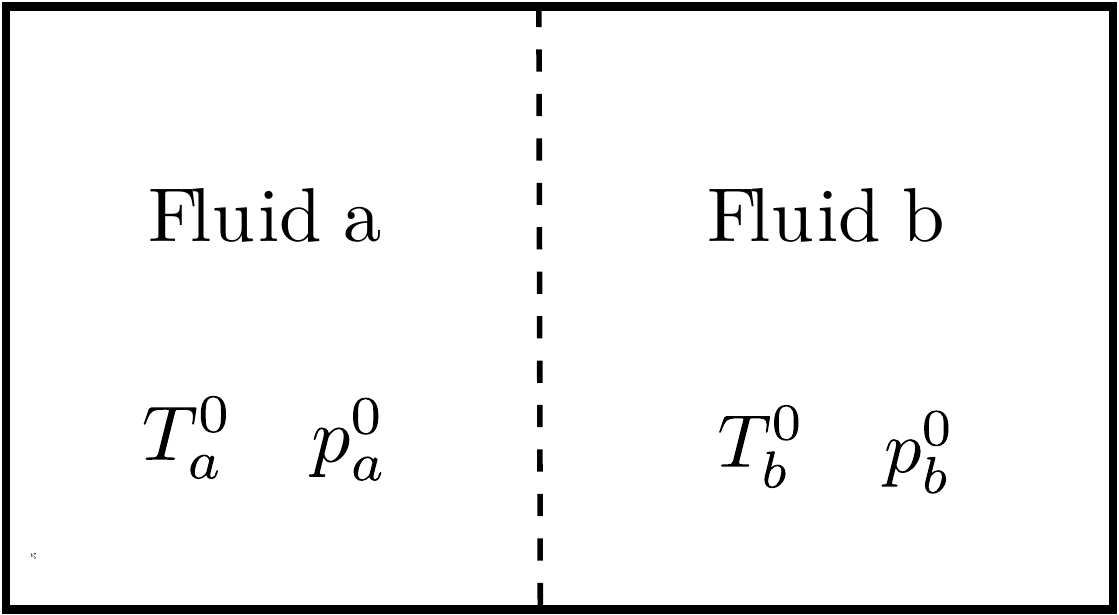}
\caption{\label{MF_EX}Two separate fluids in a cell with different temperatures}
\end{figure}
As shown in FIG.~\ref{MF_EX}, there is a cell filled with two separate fluids $a$ and $b$ with the same velocity, satisfying the initial conditions $p_a^0 = p_b ^0= p^0$ and $T_a^0 \neq T_b^0$. For the time being, we assume all that fluid variables on both sides of the material interface in the cell are constants. Resulting from the isothermal equilibrium, internal energy must exchange between these two fluids. It turns out that pressure $p_k^0$ changes to $p_k$ with fixed mass fraction $\phi_k$ and total internal energy $e$. The isothermal equilibrium is expressed as
\begin{equation*}
\frac{p_a}{\rho_a (\gamma_a-1) C_{v,a}} = T_a = T_b = \frac{p_b}{\rho_b (\gamma_b-1) C_{v,b}}.
\end{equation*}
Conservation of total internal energy before and after isothermal equilibrium shows that
\begin{equation*}
\frac{\phi_a p_a}{\rho_a(\gamma_a-1)} + \frac{\phi_b p_b}{\rho_b(\gamma_b-1)} = e = \phi_a C_{v,a} T_a^0 + \phi_b C_{v,b} T_b^0.
\end{equation*}
Then $p_a,p_b$ are known. In the end,  the pressure $p$ in the cell after isothermal equilibrium  described by Dalton's law
\begin{align*}
p=& \frac{\rho\phi_a}{\rho_a} p_a + \frac{\rho\phi_b}{\rho_b} p_b\nonumber\\
 =& p^0-\frac{\rho \phi_a\phi_b C_{v,a}C_{v,b}}{\phi_a C_{v,a}+\phi_b C_{v,b}}  (\gamma_a-\gamma_b)(T_a^0-T_b^0)
\end{align*}
is not equal to $p^0$. It turns out that a dramatic change of the pressure arises from the material interface of the fluid mixture. To prevent the pressure error, some shock-capturing schemes using energy corrections was designed for  a UPV flow\cite{banks_high-resolution_2007} or the convection of the internal energy\cite{jenny_correction_1997}. These schemes can reduce the pressure error efficiently, but their total energy looses  conservativity. Sometimes, this non-conservation destroys the simulation of internal energy distribution (see the example in Sec.~\ref{sec:int_err}) or obtains more inaccurate shock position (example in Sec.~\ref{sec:sod}).
Hence, we had better build an appropriate conservative model evading the isothermal equilibrium, i.e., internal energy exchange between two fluids, in the computational cells.

\subsection{Volume fraction model with isobaric hypothesis}

To maintain pressure equilibrium across  material interfaces, as shown in Refs.~\onlinecite{allaire_five-equation_2002,shyue_efficient_1998}, an isobaric hypothesis $p_a = p_b$ is made to express $\gamma$ in Eq.~\eqref{gamma} explicitly by 
\begin{equation}
\frac{1}{\gamma-1} = \frac{z_a}{\gamma_a-1}+\frac{z_b}{\gamma_b-1}.
\end{equation}
Since the volume fraction $z_a$ propagates with the fluid velocity $\bm{u}$, the transport equation for $z_a$ is written as
\begin{equation}\label{Z_a_eq}
\frac{\partial}{\partial t} z_a + \bm{u} \cdot \nabla z_a = 0.
\end{equation}
The use of this transport equation results in non-conservative schemes, which, though, can effectively prevent the pressure oscillations around the interfaces. For the computation of compressible multi-fluid flows, the underlying scheme is often required to be conservative to capture shocks correctly (an example can be found in Sec.~4.1 of Ref.~\onlinecite{abgrall_how_1994}, and  in Sec.~\ref{sec:sod} of the present study), for which Godunov-based schemes are a natural choice.
Therefore, we have to face on the challenge due to  the conflict between the non-conservativity for capturing interfaces and the conservativity for  capturing shocks. 

\section{\label{Sec:En-split}An energy-splitting method without internal energy exchange}

In view of the analysis in the previous section,  it is a plausible way to make isobaric hypothesis in order to design a non-oscillatory conservative scheme for multi-fluid flows.  In the following, we will explain the motivation and rationality of an energy-splitting method without internal energy exchange between materials, based on the four-equation model and Dalton's law of partial pressures.

\subsection{\label{sec:no-internal}Hypothesis: no internal energy exchange between materials}
At first, we use the mass fraction model with isobaric hypothesis. In this case, we can compute the mass fraction through Eq.~\eqref{MI_eq}. Note that this equation does not provide the volume fraction directly so that we need to close the thermodynamic system for computing the volume fraction based on the isobaric hypothesis and other reasonable physical hypothesis.
Recalling Sec.~\ref{sec:isothermal}, we have proved that there is pressure error across material interfaces due to isothermal hypothesis, which means the temperature of two materials reaching temperature equilibrium in a cell. In the process of temperature equilibrium, the exchange of the temperature between two materials in a cell causes the pressure error. Hence, it is reasonable to avoid changes of the temperature inside cells containing material interfaces, which means the temperature of two materials inside cells remain unchanged. It implies that there is no exchange of internal energy between materials.
\begin{figure}[ht]
\centering
\includegraphics[width=0.8\linewidth]{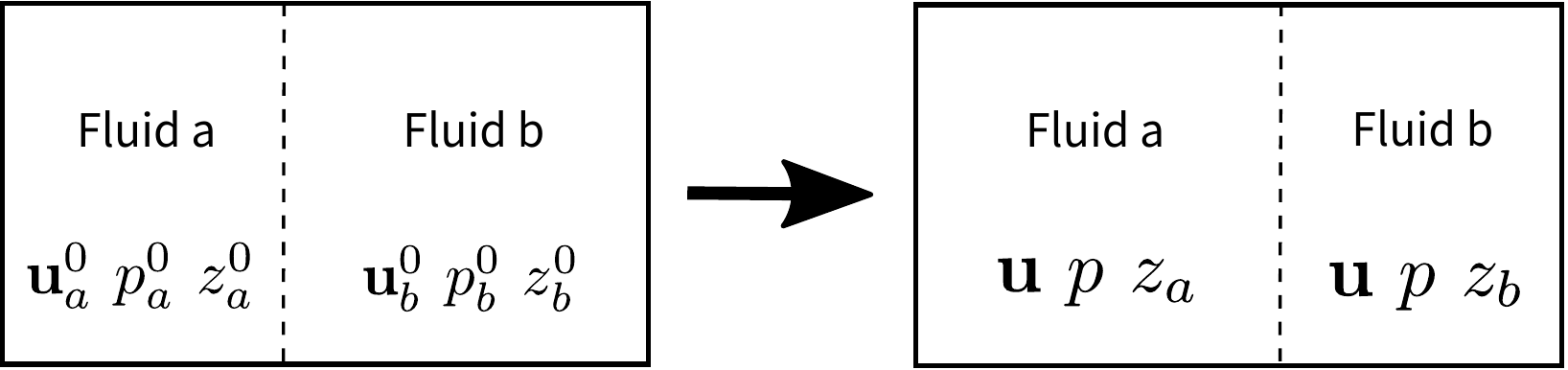}
\caption{\label{MF_EX2}Two separate fluids in a cell with different pressures}
\end{figure}
As shown in FIG.~\ref{MF_EX2}, there is a cell full of two separate fluids $a$ and $b$ moving with a common velocity $\bm{u}_a^0=\bm{u}_b^0=\bm{u}$, but with different initial pressures $p_a^0 \neq p_b ^0$. Using the Godunov averaging (first order), all fluid variables are considered as constants in the cell. The initial volume fraction of fluid $k$ is $z_k^0$. As physical quantities are in equilibrium in the cell, the velocity, pressure and the volume fraction become $\bm{u},p$ and $z_k$, respectively.  We assume that there is no exchange of internal energy between two materials in the equilibrium process. Thus after the isobaric equilibrium, density $\rho\phi_k$ and specific internal energy $e_k$ for fluid $k$ in a cell remain unchanged. As the internal energy $\rho\phi_ae_a$ and $\rho\phi_be_b$ remain unchanged
\begin{equation*}
\left\{
\begin{aligned}
\frac{z_a p}{\gamma_a-1} = \rho \phi_a e_a = \frac{z_a^0 p_a^0}{\gamma_a-1},\\
\frac{z_b p}{\gamma_b-1} = \rho \phi_b e_b = \frac{z_b^0 p_b^0}{\gamma_b-1},
\end{aligned}
\right.
\end{equation*}
we note that  Dalton's law still holds 
\begin{equation*}
p=z_a^0 p_a^0+z_b^0 p_b^0,
\end{equation*}
and find 
\begin{equation*}
z_a=z_a^0\frac{p_a^0}{z_a^0 p_a^0+z_b^0 p_b^0}\not= z_a^0,
\end{equation*}
which means that the material interface moves inside the cell. This procedure can be found in Ref.~\onlinecite{jenny_correction_1997}. Thus in the Eulerian framework, the equation \eqref{Z_a_eq} is imperfect in the case of non-equilibrium pressure between the two fluids. However, for this case, we can compute the effective ratio of specific heats for the mixture in Eq.~\eqref{gamma} through the internal energy for individual fluid
\begin{equation}
\gamma = \frac{\phi_a e_a \gamma_a + \phi_b e_b \gamma_b}{e}, 
\end{equation}
as long as we know the specific internal energy $e_k$. Then, an important question is how to compute $e_k$.  In this subsection, we only consider the two separate fluids $a$ and $b$ moving with a common velocity. So what happens to fluids at different velocities? And do the different velocities of fluids $a$ and $b$ affect the computation of $e_k$?  We will discuss these in the next subsection.

\subsection{\label{sec:kinetic}Computation of the kinetic energy exchange in a cell}

Since there is only one common velocity in the present model,  a process of velocity uniformity in the cell arises when two separate fluids with different velocities enter a same cell. As far as the  interaction of shock wave and material interface is studied, this situation must happen. After the process of velocity uniformity, the momentum of the two fluids has been exchanged, which causes the exchange of kinetic energy between fluids. However, the total momentum and total energy of two fluids remain unchanged throughout the process. Similarly as shown in FIG.~\ref{MF_EX}, we assume that there is a cell filled with two separate fluids $a$ and $b$ moving with different velocities $\bm{u}_a^0 \neq \bm{u}_b^0$.  However in practical computation, we consider the two fluids moving with a uniform velocity $\bm{u}$.  This process of uniformizing different velocities in a cell means the velocities of fluids $\bm{u}_a^0$ and $\bm{u}_b^0$ both become $\bm{u}$.  In this process, the total momentum remains unchanged
\begin{equation}
\rho\bm{u}=\rho\phi_a\bm{u}_a^0+\rho\phi_b\bm{u}_b^0.
\end{equation}
The kinetic energy of $a$ is increased by
\begin{align*}
\Delta E_{K,a} =& \frac{1}{2}\rho\phi_a|\bm{u}|^2-\frac{1}{2}\rho\phi_a |\bm{u}_a^0|^2\nonumber\\
=& \frac{1}{2}\rho\phi_a\phi_b(\bm{u}_b^0-\bm{u}_a^0)\cdot(\bm{u}+\bm{u}_a^0),
\end{align*}
and the kinetic energy of $b$ is decreased by
\begin{align*}
\Delta E_{K,b} =& \frac{1}{2}\rho\phi_b |\bm{u}_b^0|^2-\frac{1}{2}\rho\phi_b |\bm{u}|^2\nonumber\\
=& \frac{1}{2}\rho\phi_a\phi_b(\bm{u}_b^0-\bm{u}_a^0)\cdot(\bm{u}+\bm{u}_b^0),
\end{align*}
where the subscript `K' represents "Kinetic". The changes of kinetic energy for fluids $a$ and $b$ indicate the exchange of kinetic energy between fluids.
Then under  a rough hypothesis that energy exchange meets the principle of mass fraction distribution, the energy of fluid $a$ is increased from the energy of $b$ in amount of 
\begin{align}
\Delta E_K = & \phi_a\Delta E_{K,a}+\phi_b\Delta E_{K,b}\nonumber\\
= & \rho\phi_a\phi_b(\bm{u}_b^0-\bm{u}_a^0)\cdot\bm{u}\nonumber\\
= & (\rho\phi_a\bm{u}-\rho\phi_a\bm{u}_a^0)\cdot\bm{u}.
\end{align}
Using this kinetic energy exchange, we can simulate the values of  total energy for two separate fluids at each step of velocity uniformity. After the kinetic energy exchange, the total energy of fluid $a$ before equilibrium
\begin{equation*}
E_a^0=\frac{z_a^0 p_a^0}{\gamma_a-1}+\frac{1}{2}\rho\phi_a |\bm{u}_a^0|^2
\end{equation*}
becomes
\begin{equation}\label{a_kinetic_ex}
\rho\phi_a e_a+\frac{1}{2}\rho\phi_a |\bm{u}|^2 = E_a^0 +\Delta E_K.
\end{equation}
Similarly, as the total energy of two fluids remains unchanged, the initial total energy of fluid $b$ becomes 
\begin{equation}
\rho\phi_b e_b+\frac{1}{2}\rho\phi_b |\bm{u}|^2 = E_b^0 -\Delta E_K.
\end{equation}
According to the above formulae, we obtain the internal energy $e_k$.

This process of kinetic energy exchange is independent of the previous pressure equilibrium process. Without the manipulation of kinetic energy, the model can be only operated normally in the flow field with small pressure and velocity gradients across material interfaces. For other methods, only the convection of internal energy is considered, e.g. in Ref.~\onlinecite{ton_improved_1996}, without the exchange of kinetic energy. However, in the case of large velocity gradient, such as the numerical simulation of detonation process, the manipulation of kinetic energy is necessary. See Sec.~\ref{Sec:Why_ex} for the numerical evidence.

\subsection{Computation of the volume fraction for fluid $a$}

In the following, we give the partial differential equations to compute the momentum and total energy of fluid $k$.
We consider that the pressure reaches equilibrium  and the velocity reaches uniform in a flash on the entire computational region.
For a fluid parcel containing two fluids, in light of  Dalton's law with equilibrium pressure $p$, the partial pressure of fluid $k$ is $z_k^0 p$, where $z_k^0$ is the volume fraction of fluid $k$ in the fluid parcel.
 In the numerical computation, we take each computational cell as a fluid parcel. Motivated by a single-pressure compressible stratified flow model introduced in Ref.~\onlinecite{chang_robust_2007}, we assume that the partial pressures of fluids also obey Dalton's law and reach equilibrium at cell interfaces. Then, with the equilibrium pressure $p$ at cell interfaces, the partial pressure $z_a^0 p$ determines the pressure terms of fluxes in the momentum equation and energy equation for fluid $a$. And we assume the velocities for two fluids are uniform in the fluxes at cell interfaces. Then the momentum equation of fluid $a$ is
\begin{equation}\label{mo_eq_a}
\frac{\partial}{\partial t}(\rho\phi_a \bm{u}_a)+\nabla \cdot(\rho\phi_a \bm{u}\otimes\bm{u}+z_a^0 p\bm{I})=0,
\end{equation}
and the momentum equation of fluid $b$ is
\begin{equation}
\frac{\partial}{\partial t}(\rho\phi_b \bm{u}_b)+\nabla \cdot(\rho\phi_b \bm{u}\otimes\bm{u}+z_b^0 p\bm{I})=0,
\end{equation}
where $\bm{u}_k$ is the velocity of fluid $k$. In these two momentum equations, the convection term in the flux depends on the mass fraction, yet the pressure term in the flux depends on the volume fraction, which are physically reasonable. The addition of these two momentum equations is Eq.~\eqref{momentum_eq}.
According to these two momentum equations, we can obtain the velocities $\bm{u}_a$ and $\bm{u}_b$ after each computational time step, and $\bm{u}_a$ may not be equal to $\bm{u}_b$. Therefore, after each computational time step, we have a process of uniformizing different velocities in Sec.~\ref{sec:kinetic}, i.e. the kinetic energy of fluid $a$ increased from the kinetic energy of $b$ in amount of 
\begin{equation}\label{DE_k}
\Delta E_K = (\rho\phi_a\bm{u}-\rho\phi_a\bm{u}_a)\cdot\bm{u}.
\end{equation}
After that, the velocities $\bm{u}_a$ and $\bm{u}_b$ reach a uniform velocity $\bm{u}$, which is the initial velocity of $a$ and $b$ at the next computational step.

In addition, the energy equation of fluid $a$ in conservative form is
\begin{align}
&\frac{\partial}{\partial t}\left[\rho\phi_a\left(e_a+\frac{1}{2}|\bm{u}|^2\right)\right]\nonumber\\
&+\nabla \cdot\left[\rho\phi_a\bm{u}\left(e_a+\frac{1}{2}|\bm{u}|^2\right)+z_a^0 p\bm{u}\right]=0,\label{en_eq_a}
\end{align}
where $\nabla \cdot (z_a^0 p\bm{u})$ represents  pressure work. Thus the energy equation \eqref{energy_eq} can be split into two parts:  Eq.~\eqref{en_eq_a} and
\begin{align}
&\frac{\partial}{\partial t}\left[\rho\phi_b\left(e_b+\frac{1}{2}|\bm{u}|^2\right)\right]\nonumber\\
&+\nabla \cdot\left[\rho\phi_b\bm{u}\left(e_b+\frac{1}{2}|\bm{u}|^2\right)+z_b^0 p\bm{u}\right]=0.
\end{align}
According to the energy equation of fluid $a$, we can obtain the total energy of fluid $a$, $\rho\phi_a\left(e_a+\frac{1}{2}|\bm{u}|^2\right)$, after each time step. Thanks to the process of velocity uniformization, the total energy of fluid $a$ is increased by $\Delta E_K$ in Eq.~\eqref{DE_k} at each step. This is the process of the kinetic energy exchange at material interfaces described in  Eq.~\eqref{a_kinetic_ex}. 
Furthermore, according to hypothesis that $e_a,e_b$ do not exchange with each other as described in Sec.~\ref{sec:no-internal}, the pressure equilibrium process does not cause any change in specific internal energy $e_k$.
Then we can use the total energy of fluids $a$ available in Eq.~\eqref{a_kinetic_ex} to compute $e_a$ after pressure equilibrium. Based on the isobaric hypothesis, we are able to obtain the volume fraction $z_a$ by Eq.~\eqref{fact} and
\begin{equation*}
\rho_a e_a(\gamma_a-1)=p=\rho_b e_b(\gamma_b-1),
\end{equation*}
which imply
\begin{equation}\label{z_a_comp}
z_a=\frac{\rho \phi_a e_a(\gamma_a-1)}{p}.
\end{equation}
This is the initial volume fraction of fluid $a$ in cells at the next computational step.
Another form of  total energy equation for fluid $a$ in non-conservative form was proposed in Ref.~\onlinecite{miller_high-order_1996}, and  modified in Ref.~\onlinecite{ton_improved_1996} to compute partial pressures and ratios of specific heats. However, since the equation is in non-conservative form in those studies, it is difficult to define numerical integral paths for constructing a conservative finite volume scheme converging to correct weak solution\cite{abgrall_comment_2010}, especially over unstructured meshes.

\section{\label{Sec:Non-Osc}A Non-Oscillatory Conservative Scheme for Capturing Material Interfaces}

In order to maintain pressure equilibrium and mass conservation of each material across the material interface, 
we use the mass fraction model with isobaric hypothesis. Therefore, for any infinitesimal fluid parcel in which the flow field is continuously differentiable, the governing equations for the fluid mixture and the fluid $a$ take the conservative form
\begin{equation}\label{Euler_eq}
\frac{\partial }{\partial t}\bm{U}+\nabla \cdot \bm{F}(\bm{U})+\nabla \cdot \left(z_a^0 \bm{G}(\bm{U})\right)=0,
\end{equation}
with
\begin{align*}
&\bm{U}=
\begin{bmatrix}
\rho\\
\rho \bm{u}\\
\rho \left(e+\frac{1}{2} |\bm{u}|^2\right)\\
\rho \phi_a\\
\rho\phi_a \bm{u}_a\\
\rho\phi_a\left(e_a+\frac{1}{2}|\bm{u}|^2\right)
\end{bmatrix},
&\bm{G}=
\begin{bmatrix}
0\\
0\\
0\\
0\\
p\bm{I}\\
p\bm{u}
\end{bmatrix},\nonumber\\
&\bm{F}=
\begin{bmatrix}
\rho \bm{u}\\
\rho \bm{u}\otimes \bm{u} + p \bm{I}\\
\rho \bm{u}\left(e+\frac{1}{2} |\bm{u}|^2\right)+p\bm{u}\\
\rho\phi_a \bm{u}\\
\rho\phi_a \bm{u}\otimes\bm{u}\\
\rho\phi_a \bm{u}\left(e_a+\frac{1}{2}|\bm{u}|^2\right)
\end{bmatrix},&
\end{align*}
where $z_a^0$ is the volume fraction of fluid $a$ in the fluid parcel (computational cell), and $\bm{u}_a$ is the velocity of fluid $a$.
A system of equations, similar to \eqref{Euler_eq},  can be derived for fluid $b$, but both of them are equivalent. 

The conservative form of \eqref{Euler_eq} allows us to use the finite volume framework to design  numerical schemes, particularly in multi-dimensions.   

\subsection{A full-discrete finite volume method}
\begin{figure}[ht]
\centering
\includegraphics[width=0.45 \linewidth]{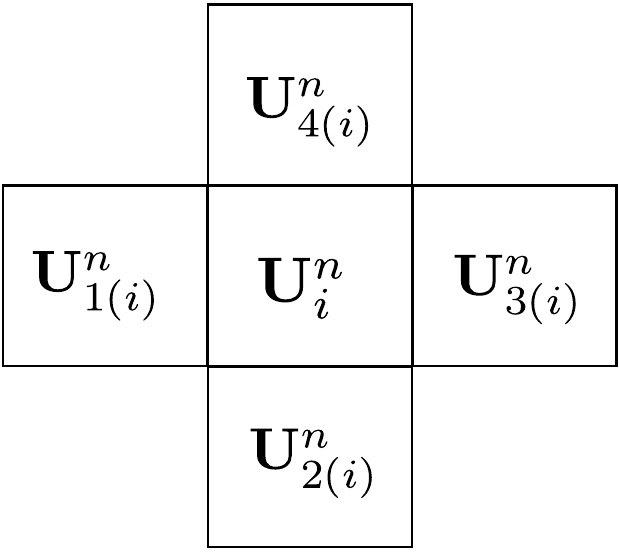}
\caption{\label{Rec}Rectangular cells and the distribution of the solution}
\end{figure}

We discretize the governing equations \eqref{Euler_eq} with a cell-centered finite-volume scheme over a two-dimensional computational domain  divided into a set of polygonal cells $\{\Omega_i\}$. The integral average of the solution vector $\bm{U}(\bm{x},t_n)$, $\bm{x} =(x,y)$,  over cell $\Omega_i$ at  time $t_n$ is given by $\bm{U}_i^n$. Taking rectangular cells as an example, we denote $\bm{U}_{j(i)}^n$ as the integral average over the $j$-th adjacent cell  $\Omega_{j(i)}$ of $\Omega_i$, as in FIG.~\ref{Rec}.  
The Godunov scheme assumes that the fluid data at  time $t_n$ are piece-wise constant distribution.
Then taking the cell as a fluid parcel, we know that the volume fraction in the cell $\Omega_i$ is $z_a^0=z_{a,i}^n$. By solving the exact Riemann problem $\mbox{RP}\left(\bm{U}_{j(i)}^n,\bm{U}_i^n\right)$, we can obtain the solution $\bm{U}_{i,j}^n$ at the $j$-th boundary between $\Omega_i$ and $\Omega_{j(i)}$. Then the finite-volume scheme with the Godunov fluxes is given by
\begin{equation}
\bm{U}_i^{n+1} =  \bm{U}_i^{n} - \sum_{j=1}^4 \Lambda_j^i\left[\bm{H}_j\left(\bm{U}_{i,j}^n\right)+z_{a,i}^n\bm{K}_j\left(\bm{U}_{i,j}^n\right)\right],\label{FV}
\end{equation}
where $\Lambda_j^i=\Delta t
\, L_j/|\Omega_i|$, $L_j$ is the length of the $j$-th boundary of the cell $\Omega_i$, $|\Omega_i|$ is the volume of $\Omega_i$, $\bm{n}_j$ is the unit vector outward normal to the $j$-th boundary, $\bm{H}_j=\bm{F}\cdot \bm{n}_j$ and $\bm{K}_j=\bm{G}\cdot \bm{n}_j$. Specifically, the exact Riemann problem $\mbox{RP}\left(\bm{U}_{j(i)}^n,\bm{U}_i^n\right)$ is solved for  the planar one-dimensional Euler equations in the normal direction  $\bm n_j$  of  the boundary between $\Omega_i$ and $\Omega_{j(i)}$, 
\begin{equation}\label{1D-Euler}
\frac{\partial }{\partial t}\bm{U}+\frac{\partial }{\partial \bm{n}_j}\bm{H}_j=0. 
\end{equation}
The effective ratio of specific heats in $\Omega_i$ is computed by using Eq.~\eqref{gamma_mix}
\begin{equation}
\gamma_i^n=\frac{(\rho \phi_a e_a)_i^n \gamma_a + \left[(\rho  e)_i^n-(\rho \phi_a e_a)_i^n\right] \gamma_b}{(\rho e)_i^n}.
\end{equation}
Detailed process of the computation for the Godunov fluxes can be found in Refs.~\onlinecite{ben-artzi_generalized_2003} or \onlinecite{toro_riemann_1997}. Using the solution $\bm{U}_{i,j}^n$ at the $j$-th boundary, the mass fraction and volume fraction in the fluxes are determined as
\begin{equation}
\phi_{a,i,j}^n=\left\{
\begin{aligned}
&\phi_{a,i}^n&\mbox{if }\bm{u}_{i,j}^n\cdot \bm{n}_j>0,\\
&\phi_{a,j(i)}^n&\mbox{otherwise,}
\end{aligned}
\right.
\end{equation}
and
\begin{equation} 
z_{a,i,j}^n=\left\{
\begin{aligned}
&z_{a,i}^n & \mbox{if }\bm{u}_{i,j}^n\cdot \bm{n}_j>0,\\
&z_{a,j(i)}^n & \mbox{otherwise.}
\end{aligned}
\right.
\end{equation}
It can be proved that this condition preserves the positivity of the mass fractions\cite{larrouturou_how_1991}. Then, we have  the solution $\bm{U}_{i,j}^n$ and the fluxes at the $j$-th boundary. Especially, the last component of $\bm{H}_j(\bm{U}_{i,j}^n)$ is
\begin{align*}
&\left((z_a\rho_a e_a)_{i,j}^n+\frac{1}{2}\rho_{i,j}^n\phi_{a,i,j}^n|\bm{u}_{i,j}^n|^2\right)\bm{u}_{i,j}^n\cdot \bm{n}_j\nonumber\\
=&\left(z_{a,i,j}^n\frac{p_{i,j}^n}{\gamma_a-1}+\frac{1}{2}\rho_{i,j}^n\phi_{a,i,j}^n|\bm{u}_{i,j}^n|^2\right)\bm{u}_{i,j}^n\cdot \bm{n}_j.
\end{align*}

\subsection{Computation of internal energy for fluid $a$}
We use  the energy equation of fluid $a$ in Eqs.~\eqref{FV} to update the total energy of fluid $a$, $\rho\phi_a\left(e_a+\frac{1}{2}|\bm{u}|^2\right)$. According to the momentum equations of fluid $a$ in Eqs.~\eqref{FV}, we are able to compute the momentum for fluid $a$, $\rho\phi_a\bm{u}_a$, before the velocity uniformity at next time $t_{n+1}$. If the two immiscible fluids $a$ and $b$ have different velocities, their velocities achieve uniformity
\begin{equation*}
\bm{u}_i^{n+1}=\phi_{a,i}^{n+1}\bm{u}_{a,i}^{n+1}+\phi_{b,i}^{n+1}\bm{u}_{b,i}^{n+1}.
\end{equation*}
In the process of velocity uniformization, the exchange of  kinetic energy from fluid $b$ to fluid $a$ can be estimated as 
\begin{equation*}
(\Delta E_K)_i^{n+1} = \left((\rho\phi_a)_i^{n+1}\bm{u}_i^{n+1}-(\rho\phi_a\bm{u}_a)_i^{n+1}\right)\cdot\bm{u}_i^{n+1}.
\end{equation*}
Then we obtain the internal energy for fluid $a$ at time $t_{n+1}$, $(\rho\phi_a e_a)_i^{n+1}$, through the total energy of fluid $a$ after the velocity uniformity
\begin{align}
&(\rho\phi_a e_a)_i^{n+1}+\frac{1}{2}(\rho\phi_a)_i^{n+1}\left|\bm{u}_i^{n+1}\right|^2\nonumber\\
=&\left[\rho\phi_a\left(e_a+\frac{1}{2}|\bm{u}|^2\right)\right]_i^{n+1}+(\Delta E_K)_i^{n+1}.
\end{align}
The velocity of fluid $a$ becomes $\bm{u}_i^{n+1}$. The exchange process of kinetic energy is necessary for  extreme situations with large velocity gradient. For more extreme situations, due to the admissible error of simulating kinetic energy exchange, the internal energy of $a$ or $b$, $\rho \phi_a e_a$ or $\rho \phi_b e_b=\rho e - \rho \phi_a e_a$ may be less than 0. A computational process is to truncate it to be zero, which means the internal energy is extremely small. This situation basically does not appear with the exchange process of kinetic energy (see Sec.~\ref{Sec:Why_ex}). In this way, we can guarantee  the positivity of volume fractions, since now the volume fraction $z_a$ in $\Omega_i$ is
\begin{equation}\label{Z_a_update}
z_{a,i}^{n+1}=\frac{(\rho \phi_a e_a)_i^{n+1}(\gamma_a-1)}{(\rho \phi_a e_a)_i^{n+1}(\gamma_a-1)+(\rho \phi_b e_b)_i^{n+1}(\gamma_b-1)}.
\end{equation}
To sum up, we get a non-oscillatory Godunov scheme for two-dimensional multi-fluid flows. Such a scheme is termed as the energy-splitting  Godunov scheme ({\em ES-Godunov} for short).

\subsection{Second-order accurate extension}

We make a second-order accurate extension of ES-Goduov  by using  the generalized Riemann problem (GRP) solver  \cite{ben-artzi_direct_2006,ben-artzi_generalized_2003} ({\em ES-GRP} for short). The two-dimensional finite-volume GRP scheme for Eqs.~\eqref{Euler_eq}, is written as
\begin{equation}
\bm{U}_i^{n+1} =  \bm{U}_i^{n} - \sum_{j=1}^4 \Lambda_j^i \left[\bm{H}_j\left(\bm{U}_{i,j}^{n+\frac{1}{2}}\right)+z_{a,i}^{n+\frac{1}{2}}\bm{I}_j\left(\bm{U}_{i,j}^{n+\frac{1}{2}}\right)\right],\label{GRP_scheme}
\end{equation}
where the mid-point value $\bm{U}_{i,j}^{n+\frac{1}{2}}$ is derived analytically by resolving the GRP at the $j$-th boundary with accuracy of second order. In each cell, we project conservative variables in the form 
\begin{equation}
\bm{U}_i^n(\bm{x})=\bm{U}_i^n+\bm{\sigma}_i^n(\bm{x}-\bm{x}_i), 
\label{linear-data}
\end{equation}
where  $\bm{\sigma}_i^n$ is the gradient of solution inside the cell $\Omega_i$ at time $t=t_n$, and $\bm{x}_i$ is the centroid of $\Omega_i$.  We indicate a parameter $\alpha\in[0,2)$ and the minmod function\citep{van_leer_towards_1979,ben-artzi_direct_2006}
\begin{equation}
\Psi(a,b,c)=\left\{
\begin{aligned}
&\min(|a|,|b|,|c|) & \mbox{if } a,b,c>0,\\
&-\min(|a|,|b|,|c|) & \mbox{if } a,b,c<0,\\
&0 & \mbox{otherwise.}
\end{aligned}
\right.
\end{equation}
For the rectangular cells, the gradient
\begin{equation}
\bm{\sigma}_i^{n}=(\bm{U}_x,\bm{U}_y)_i^n
\end{equation}
is calculated as
\begin{equation*}
(\bm{U}_x)_i^n=\Psi\left(\alpha\frac{\bm{U}_{3(i)}^n-\bm{U}_i^n}{\Delta x},(\bm{U}_x)_i^{n,-},\alpha\frac{\bm{U}_i^n-\bm{U}_{1(i)}^n}{\Delta x}\right),
\end{equation*}
and
\begin{equation*}
(\bm{U}_y)_i^n=\Psi\left(\alpha\frac{\bm{U}_{4(i)}^n-\bm{U}_i^n}{\Delta y},(\bm{U}_y)_i^{n,-},\alpha\frac{\bm{U}_i^n-\bm{U}_{2(i)}^n}{\Delta y}\right),
\end{equation*}
where 
\begin{equation*}
(\bm{U}_x,\bm{U}_y)_i^{n,-}:= \left(\frac{\bm{U}_{i,3}^n-\bm{U}_{i,1}^n}{\Delta x},\frac{\bm{U}_{i,4}^n-\bm{U}_{i,2}^n}{\Delta y}\right),
\end{equation*}
and
\begin{equation}
\bm{U}_{i,j}^n=\bm{U}_{i,j}^{n-1}+\Delta t\left(\frac{\partial \bm{U}}{\partial t}\right)_{i,j}^{n-1}.
\end{equation}
We solve the generalized Riemann problem at the $j$-th boundary $\mbox{GRP}\left(\bm{U}_{j(i)}^n(\bm{x}_{i,j}),\bm{\sigma}_{j(i)}^n;\bm{U}_i^n(\bm{x}_{i,j}),\bm{\sigma}_i^n\right)$ for the planar one-dimensional Euler equations in \eqref{1D-Euler} at the center $\bm{x}_{i,j}$ to define the Riemann solution $\bm{U}_{i,j}^n$ and determine the temporal derivative $\displaystyle\left(\frac{\partial \bm{U}}{\partial t}\right)_{i,j}^n$.
The temporal variation of the fraction $z_a$ is calculated as 	
\begin{equation}
\left(\frac{\partial z_a}{\partial t}\right)_{i,j}^n=\left\{
\begin{aligned}
&-\bm{u}_{i,j}^n\cdot (\sigma_{z_a})_i^n & \mbox{if }\bm{u}_{i,j}^n\cdot \bm{n}_j>0\\
&-\bm{u}_{i,j}^n\cdot (\sigma_{z_a})_{j(i)}^n & \mbox{otherwise,}\\
\end{aligned}
\right.
\end{equation}
by adopting the equation \eqref{Z_a_eq}, where  $(\sigma_{z_a})_i^n$ is the gradient of $z_a$ inside the cell calculated by the same process as for  $\bm{\sigma}_i^n$. Thus we have the mid-point value inside cells
\begin{equation}
z_{a,i}^{n+\frac{1}{2}}=z_{a,i}^n-\frac{\Delta t}{2} \bm{u}_i^n\cdot (\sigma_{z_a})_i^n,
\end{equation}
and the mid-point values on cell interfaces
\begin{equation}
\bm{U}_{i,j}^{n+\frac{1}{2}}=\bm{U}_{i,j}^n+\frac{\Delta t}{2}\left(\frac{\partial \bm{U}}{\partial t}\right)_{i,j}^n, \end{equation}
\begin{equation}
z_{a,i,j}^{n+\frac{1}{2}}=z_{a,i,j}^n+\frac{\Delta t}{2}\left(\frac{\partial z_a}{\partial t}\right)_{i,j}^n,
\end{equation}
which further gives $\gamma_{i,j}^{n+\frac 12}$ by 
\begin{equation}
\frac{1}{\gamma_{i,j}^{n+\frac{1}{2}}-1}=\frac{z_{a,i,j}^{n+\frac{1}{2}}}{\gamma_a-1}+\frac{1-z_{a,i,j}^{n+\frac{1}{2}}}{\gamma_b-1}.
\end{equation}
Thus the last component of $\bm{H}_j(\bm{U}_{i,j}^{n+\frac{1}{2}})$ in \eqref{GRP_scheme}  becomes
\begin{align*}
&\left((z_a \rho_a e_a)_{i,j}^{n+\frac{1}{2}}+\frac{1}{2}\rho_{i,j}^{n+\frac{1}{2}}\phi_{a,i,j}^{n+\frac{1}{2}}\left|\bm{u}_{i,j}^{n+\frac{1}{2}}\right|^2\right)\bm{u}_{i,j}^{n+\frac{1}{2}}\cdot \bm{n}_j\nonumber\\
=&\left(\frac{z_{a,i,j}^{n+\frac{1}{2}}p_{i,j}^{n+\frac{1}{2}}}{\gamma_a-1}+\frac{\rho_{i,j}^{n+\frac{1}{2}}\phi_{a,i,j}^{n+\frac{1}{2}}}{2}\left|\bm{u}_{i,j}^{n+\frac{1}{2}}\right|^2\right)\bm{u}_{i,j}^{n+\frac{1}{2}}\cdot \bm{n}_j.
\end{align*}
Using the same approach as in ES-Godunov to compute the internal energy for fluid $a$, we obtain a second-order non-oscillatory scheme for two-dimensional multi-fluid flows.  This method has the non-oscillatory property.\\
 
	\noindent \textbf{Remark.} For high-order Godunov-type methods, the energy modification can be made to the averaged flow quantities, in the same way as the corresponding first order schemes.  Then we implement the linear reconstruction process to obtain the linear distribution as in \eqref{linear-data}.    

\vspace{0.2cm}

Indeed, we consider the solution of the GRP scheme at an interface with uniform velocity $\bm{u}$ and pressure $p$. Using the fact 
\begin{equation}
\sum_{j=1}^4 \Lambda_j^i \bm{n}_j = \mathbf{0},
\end{equation}
and full-discrete scheme \eqref{GRP_scheme}, we have
\begin{equation*}
(\rho\phi_a e_a)_i^{n+1} = (\rho\phi_a e_a)_i^{n} - \sum_{j=1}^4 \Lambda_j^i \left(z_{a,i,j}^{n+\frac{1}{2}}\frac{p}{\gamma_a-1}\bm{u}\cdot \bm{n}_j\right),
\end{equation*}
and
\begin{equation*}
(\rho e)_i^{n+1} =  (\rho e)_i^{n} - \sum_{j=1}^4  \Lambda_j^i \left(\frac{p}{\gamma_{i,j}^{n+\frac{1}{2}}-1}\bm{u}\cdot \bm{n}_j\right).
\end{equation*}
In view of Eq.~\eqref{Z_a_update},  we have
\begin{equation}
z_{a,i}^{n+1}\frac{p_i^{n+1}}{\gamma_a-1}=(\rho \phi_a e_a)_i^{n+1}
\end{equation}
and then 
\begin{equation}
z_{b,i}^{n+1}\frac{p_i^{n+1}}{\gamma_b-1}=(\rho e)_i^{n+1}-(\rho \phi_a e_a)_i^{n+1}, 
\end{equation}
due to the pressure equilibrium. 
Then we proceed to obtain 
\begin{equation*}
z_{a,i}^{n+1}\frac{p_i^{n+1}}{\gamma_a-1} = z_{a,i}^n\frac{ p}{\gamma_a-1} - \sum_{j=1}^4 \Lambda_j^i \left(z_{a,i,j}^{n+\frac{1}{2}}\frac{p}{\gamma_a-1}\bm{u}\cdot \bm{n}_j\right),
\end{equation*}
and
\begin{equation*}
z_{b,i}^{n+1}\frac{p_i^{n+1}}{\gamma_b-1} = z_{b,i}^n\frac{ p}{\gamma_b-1} - \sum_{j=1}^4 \Lambda_j^i \left(z_{b,i,j}^{n+\frac{1}{2}}\frac{p}{\gamma_b-1}\bm{u}\cdot \bm{n}_j\right).
\end{equation*}
Finally, we can get the pressure at time $t_{n+1}$
\begin{align*}
p_i^{n+1}=&z_{a,i}^{n+1}p_i^{n+1}+z_{b,i}^{n+1}p_i^{n+1}\nonumber\\
=&p - \sum_{j=1}^4 \Lambda_j^i(p\bm{u}\cdot \bm{n}_j)=p.
\end{align*}
This shows that no pressure oscillation  appears at the material interface.

Finally, we would like to remark that the GRP solver has the feature that the thermodynamics is deeply characterized and embedded into the scheme\cite{li_thermodynamical_2017}.

\section{\label{Sec:num_res}Numerical Results}

We present some  numerical results by using the current energy-splitting Godunov-type methods  in Section \ref{Sec:Non-Osc} .  The results are compared with those computed by the Godunov-type methods in Ref.~\onlinecite{banks_high-resolution_2007}.   We abbreviate {\em Is-Godunov} for the results by using the Godunov method with the isothermal hypothesis\cite{larrouturou_how_1991,quirk_dynamics_1996}, {\em UPV-Godunov} for the Godunov results with the energy correction based on a UPV flow\cite{banks_high-resolution_2007}, in addition to  the abbreviations:  {\em ES-Godunov} and {\em ES-GRP}.  The process of the kinetic energy exchange is added into ES-Godunov and ES-GRP for the simulation.
The following examples show that nonphysical oscillations arising from the interface are avoided by using the current non-oscillatory conservative schemes. Through the comparison with the corresponding  physical experimental results,  the numerical results show that the current schemes perform well for two-dimensional cases with very sharp interfaces. For all examples, the CFL number is taken to be 0.45. 

\subsection{\label{Sec:Why_ex} Demonstration  for the kinetic energy exchange}

\begin{table*}[htp]
\caption{\label{why_KE}The interfacial volume fraction $z_a$  at advancing time steps for the demonstration of the necessity of   kinetic energy exchange during fluid mixture.}
\begin{ruledtabular}
\begin{tabular}{cddddd}
Scheme& \multicolumn{1}{r} \mbox{Step} 1 & \multicolumn{1}{r} \mbox{Step} 2 & \multicolumn{1}{r} \mbox{Step} 3 & \multicolumn{1}{r} \mbox{Step} 4 & \multicolumn{1}{r} \mbox{Step} 5 \\[0.5mm]
\hline\\[-2mm]
ES-Godunov(NO-KE)& -0.09652& -0.07664& -0.05920& -0.04542& -0.03496\\
ES-GRP(NO-KE)& -0.09652& -0.07707& -0.05412& -0.03272& -0.02203
\\
ES-Godunov& 0.96143& 0.90025& 0.80975& 0.68928& 0.54981
\\
ES-GRP& 0.96143& 0.88862& 0.71710& 0.45799& 0.19412
\end{tabular}
\end{ruledtabular}
\vspace{0.2cm} 
\end{table*} 

The exchange of  kinetic energy in the process of fluid mixture has not been  well studied in literatures.  Hence we propose an example to show the influence  of the kinetic energy exchange on the distribution of fluids inside mixed cells. We consider an inward two-fluid compression problem, for which the initial discontinuity  at $x=0.12$ separates fluid $a$ with $\gamma_a=1.4$  in the left from fluid $b$ with $\gamma_b=3.0$ in the right. 
These two fluids  can be regarded as air in the left and wolfram in the right.
The initial data in the entire computational domain $[0,0.15]$,  composed of $250$ cells, are given as
\begin{equation*}
\setlength\arraycolsep{0pt}
\begin{array}{lclclclr}
(\rho,u,p,\phi_a)=  (&0.00129&,&\, 0&,&\, 1.01325&,1),  &~ x<0.12,\\
(\rho,u,p,\phi_b)=  (&19.237&, &-200&, &1.01325&,1),&~ x> 0.12.
\end{array}
\end{equation*}
The left boundary is a solid wall and the right boundary has an inflow condition. This problem has  exceedingly huge density ratio and velocity gradient.
We use \textit{NO-KE} to represent  no kinetic energy exchange in the scheme, and list the numerical results of $z_a$ in the $199$-th cell at time steps $1$ to $5$ in Table \ref{why_KE}. It is observed that without the process of kinetic energy exchange, the volume fraction of air at the interface  becomes negative value, which immediately ruins the numerical simulation. This shows the necessity of the  numerical correction of the kinetic energy exchange into the current  method.

\subsection{\label{sec:sod}Two-fluid shock-tube problem}

We consider a two-fluid shock-tube problem in Ref.~\onlinecite{abgrall_how_1994}. The discontinuity initially at $x=0.3$ separates air with $\gamma_a=1.4,C_{v,a}=0.72$  in the left from helium with $\gamma_b=1.67,C_{v,b}=3.11$ in the right. Then the initial data in the entire computational domain $[0,1]$,  composed of $100$ cells,  are given by
\begin{equation*}
\setlength\arraycolsep{0pt}
\begin{array}{lclll}
(\rho,u,p,\phi_a) = (&1&,0,25,1), &~ x<0.3,\\
(\rho,u,p,\phi_b) = (&0.01&,0,20,1), &~ x > 0.3.
\end{array}
\end{equation*}
The exact solution of the shock-tube problem consists of a left-propagating rarefaction wave, a contact discontinuity moving at the speed of $0.83$, and a right-propagating shock wave at the speed of $58.35$. We compare the  solutions computed by  different schemes at time $t=0.008$.
\begin{figure*}[htb]
\begin{minipage}[t]{0.4\linewidth}
\centering
\includegraphics[width=\textwidth]{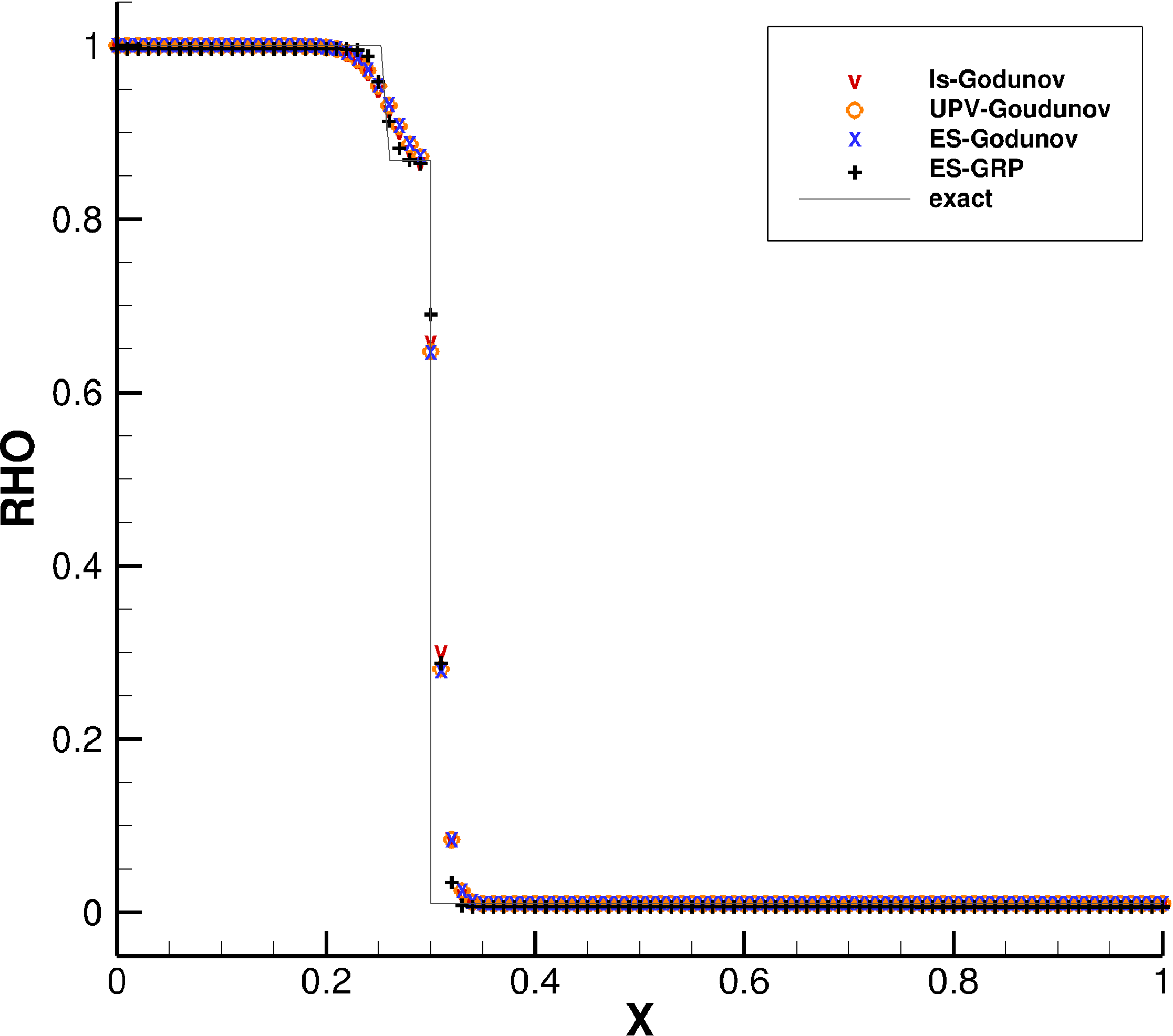}
(a)density
\end{minipage}
\hfill
\begin{minipage}[t]{0.4\linewidth}
\centering
\includegraphics[width=\textwidth]{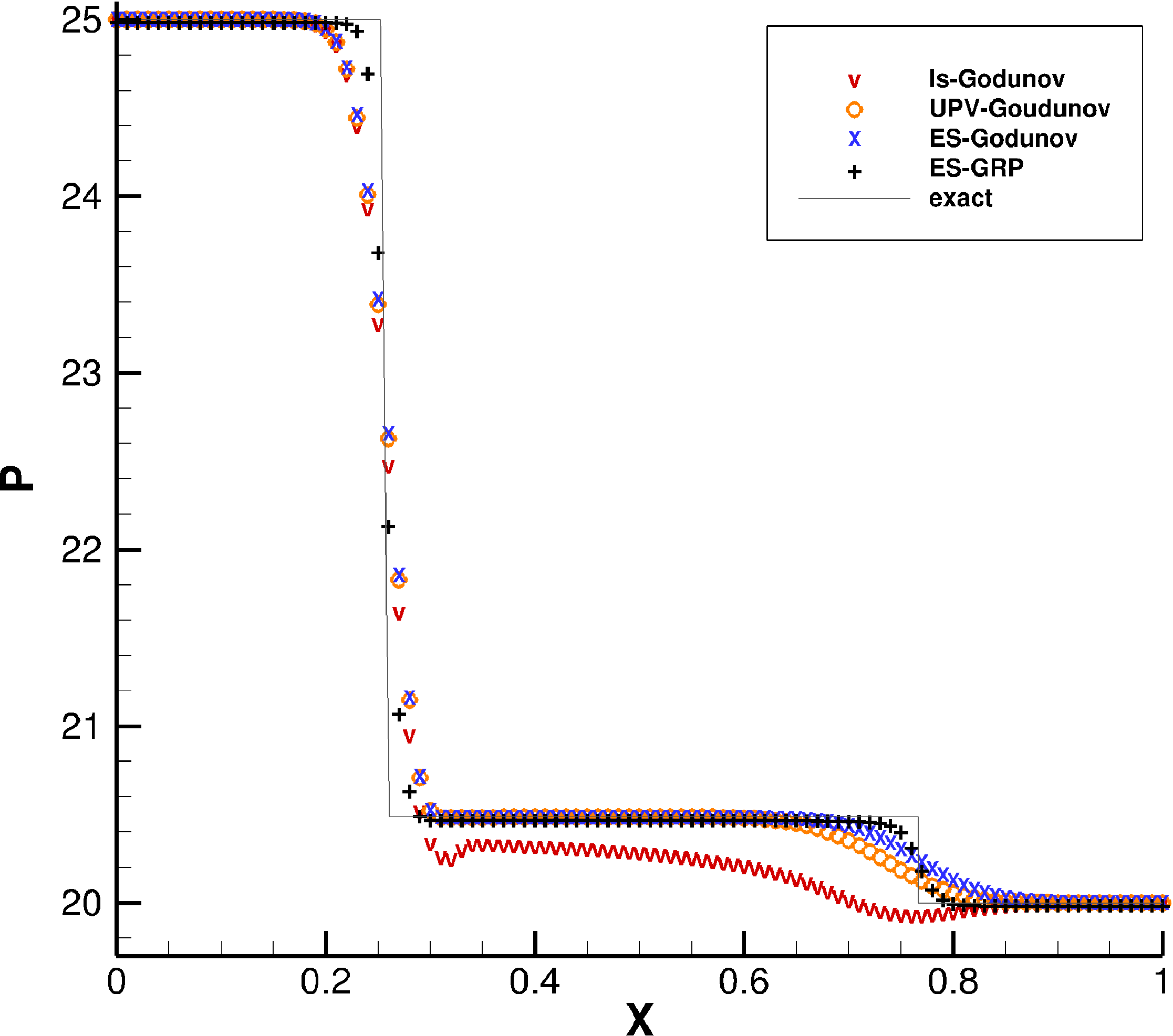}
(b)pressure
\end{minipage}
\begin{minipage}[t]{0.4\linewidth}
\centering
\includegraphics[width=\textwidth]{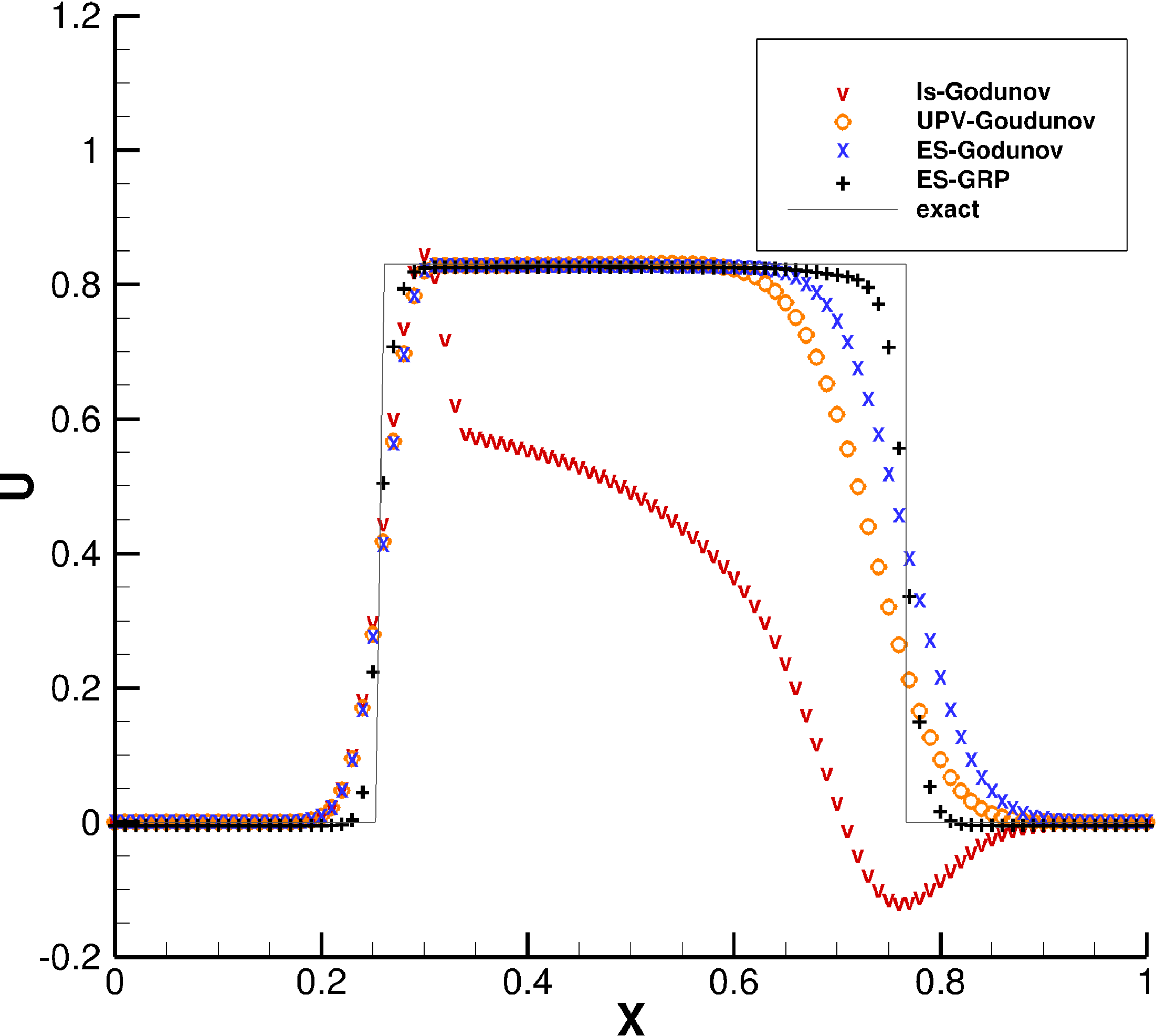}
(c)velocity
\end{minipage}
\hfill
\begin{minipage}[t]{0.4\linewidth}
\centering
\includegraphics[width=\textwidth]{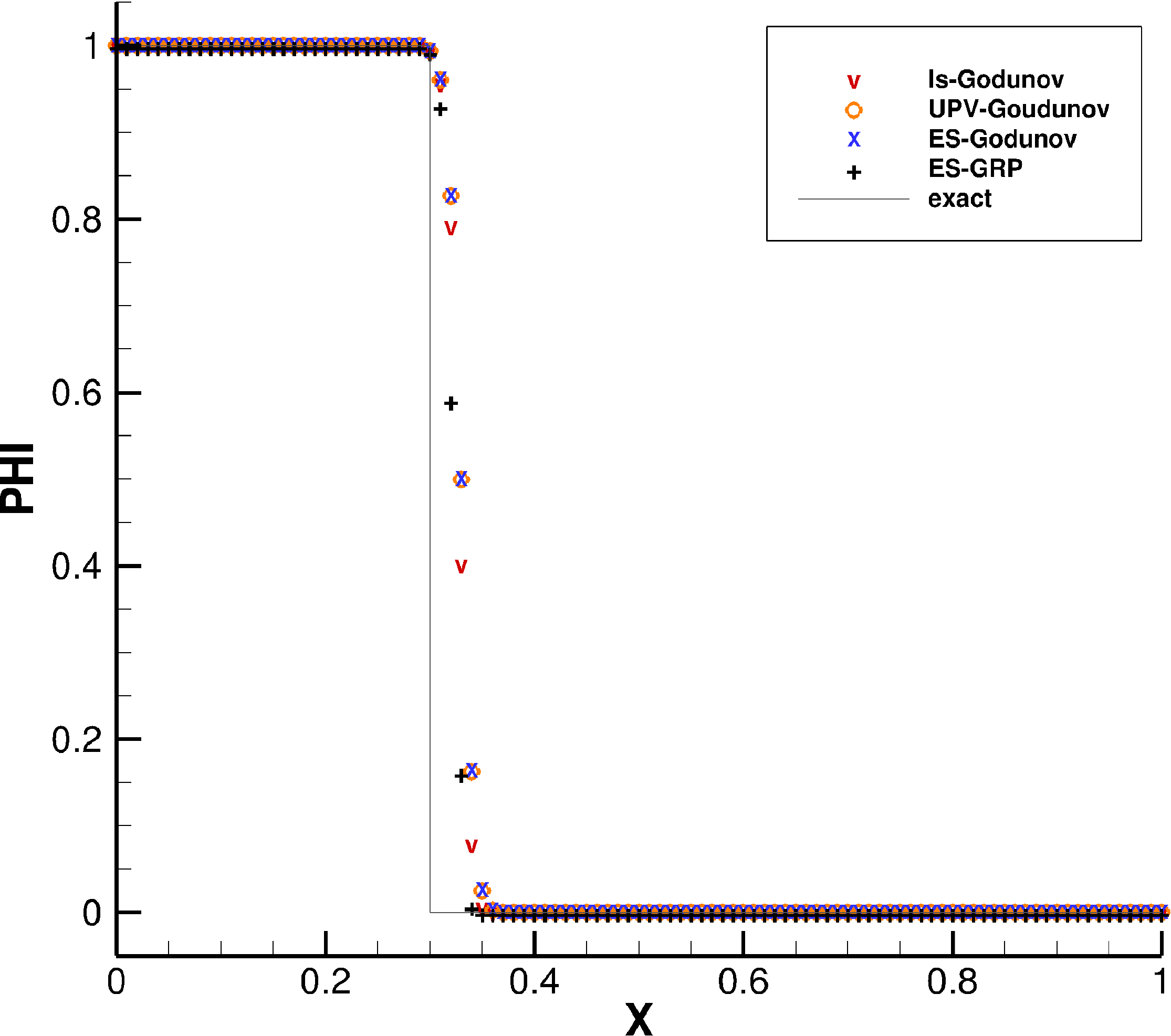}
(d)mass fraction of fluid $a$
\end{minipage}
\caption{\label{Sod}Results of the two-fluid Sod problem at $t=0.008$}
\end{figure*} 

All numerical solutions with the mass fraction model are shown in FIG \ref{Sod}. The gray curves are the exact solution; the red marks ``v'' are the solution by Is-Godunov; the orange circles are the solution by UVP-Godunov; the blue marks ``x'' are the solution by the current method ES-Godunov, and  the black plus signs represent the solution by ES-GRP with $\alpha=1.5$. The numerical  errors of pressure and velocity occur in the Is-Godunov solution.   The results by the current ES-Godunov  and ES-GRP are  much closer to the exact solution without oscillations than that by UPV-Godunov. This shows the performance of the current schemes.

\subsection{\label{sec:int_err}Shock-interface interaction}
\begin{figure*}[htb]
\begin{minipage}[t]{0.4\linewidth}
\centering
\includegraphics[width=\textwidth]{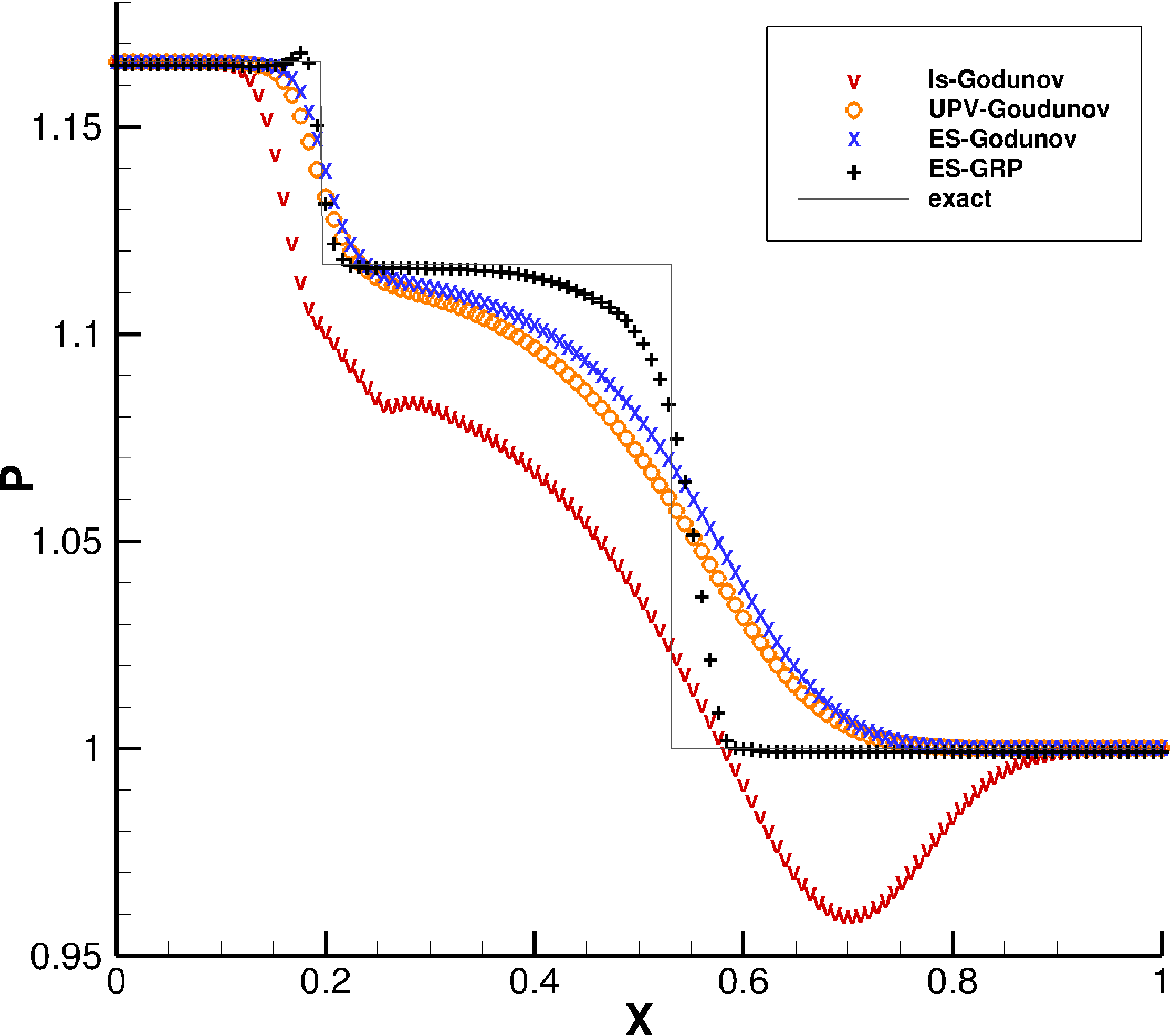}
(a)pressure
\end{minipage}
\hfill
\begin{minipage}[t]{0.4\linewidth}
\centering
\includegraphics[width=\textwidth]{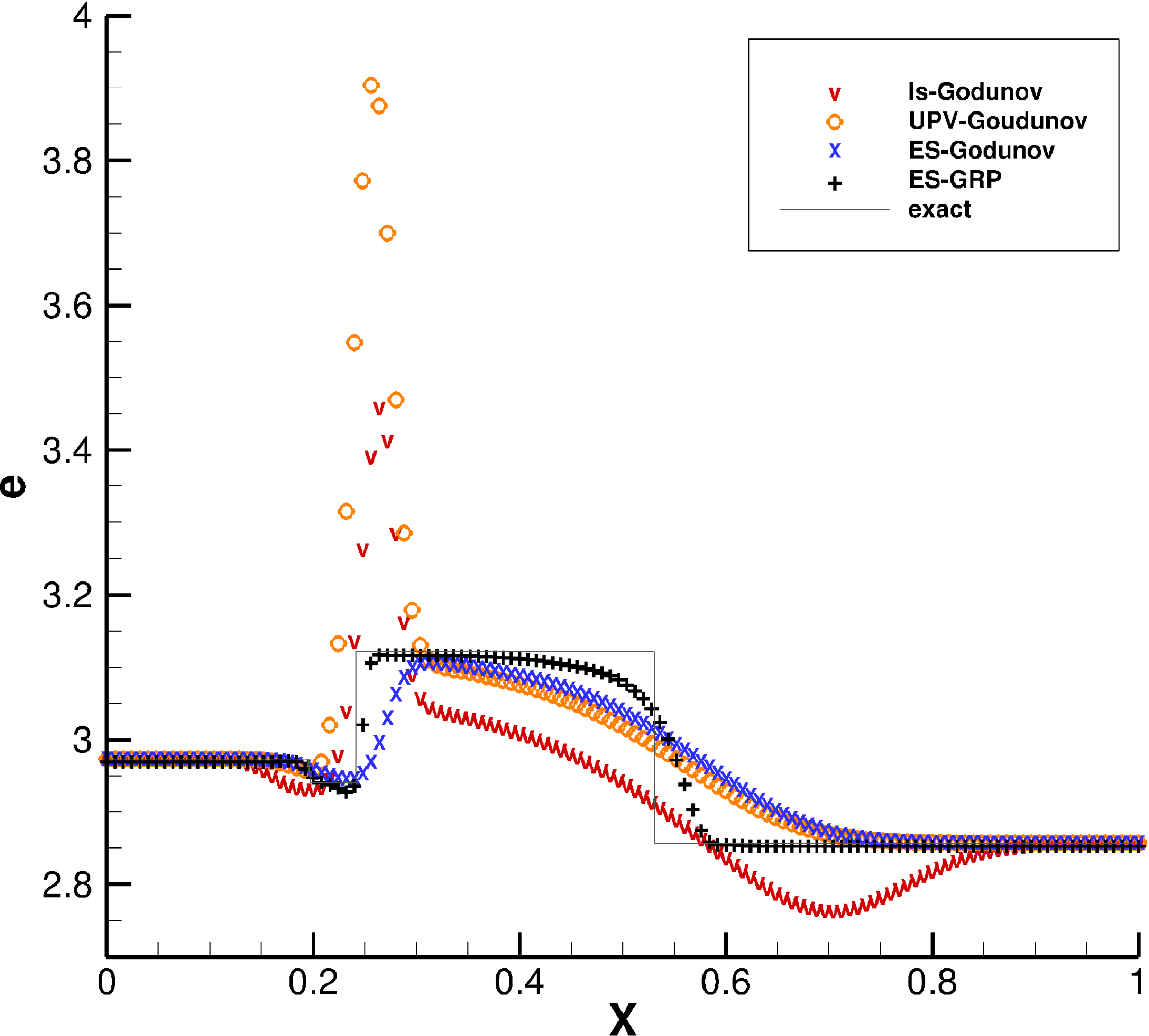}
(b)specific internal energy
\end{minipage}
\caption{\label{SIM}Results of the shock-interface interaction problem at $t=0.07$.}
\end{figure*}

We consider a shock-interface interaction problem.  The interface initially at $x=0.2$ separates fluid $a$  with $\gamma_a=1.35,C_{v,a}=2.4$ in the left from fluid $b$ with $\gamma_b=5.0,C_{v,b}=1.5$ in the right. These two materials, used in Ref.~\onlinecite{banks_high-resolution_2007}, correspond to high explosive products in the left and a confining material in the right. The interface and a shock wave with the shock Mach number $M_s=1.5$ initially at $x=0.16$ propagate to the right at the speed of $0.5$ and $1.74$, respectively. Then the initial data in the computational domain $[0,1]$,  composed of $125$ cells,  are given by
\begin{equation*}
\setlength\arraycolsep{0pt}
\begin{array}{lclclclr}
(\rho,u,p,\phi_a)=  (&1.1201&,&\, 0.6333&,&\, 1.1657&,1),  &~ x<0.16,\\
(\rho,u,p,\phi_a)=  (&1&, &0.5&, &1&,1),  &~ 0.16< x<0.2,\\
(\rho,u,p,\phi_b)=  (&0.0875&, &0.5&, &1&,1),&~ x> 0.2.
\end{array}
\end{equation*}
At  time $t=0.0322$, the interface is impacted by the shock wave. The resulting wave pattern after the interaction consists of a reflected rarefaction wave, an interface at the speed of $0.67$, and a transmitted shock at the speed of $8.32$. We compare the profiles of pressure and internal energy by using different methods at $t=0.07$ in FIG.~\ref{SIM}. The parameter of the GRP method is $\alpha=1.9$ for this example.
\vspace{0.2cm} 

Each mark represents the same solution as in the previous example. Serious pressure oscillations are generated from the interface and induced error of the internal energy occur in Is-Godunov or UVP-Godunov solutions. Although UVP-Godunov scheme  can prevent the pressure oscillations, incorrect  internal energy shows the defect of such a scheme. Therefore it is reasonable to  believe that other methods with energy correction, such as that in Ref.~\onlinecite{jenny_correction_1997}, may obtain incorrect numerical results of internal energy. In contrast, the current method can produce much better results.

\subsection{Shock-bubble interactions}
 
\begin{figure}[ht]
\centering
\includegraphics[width=0.45\textwidth]{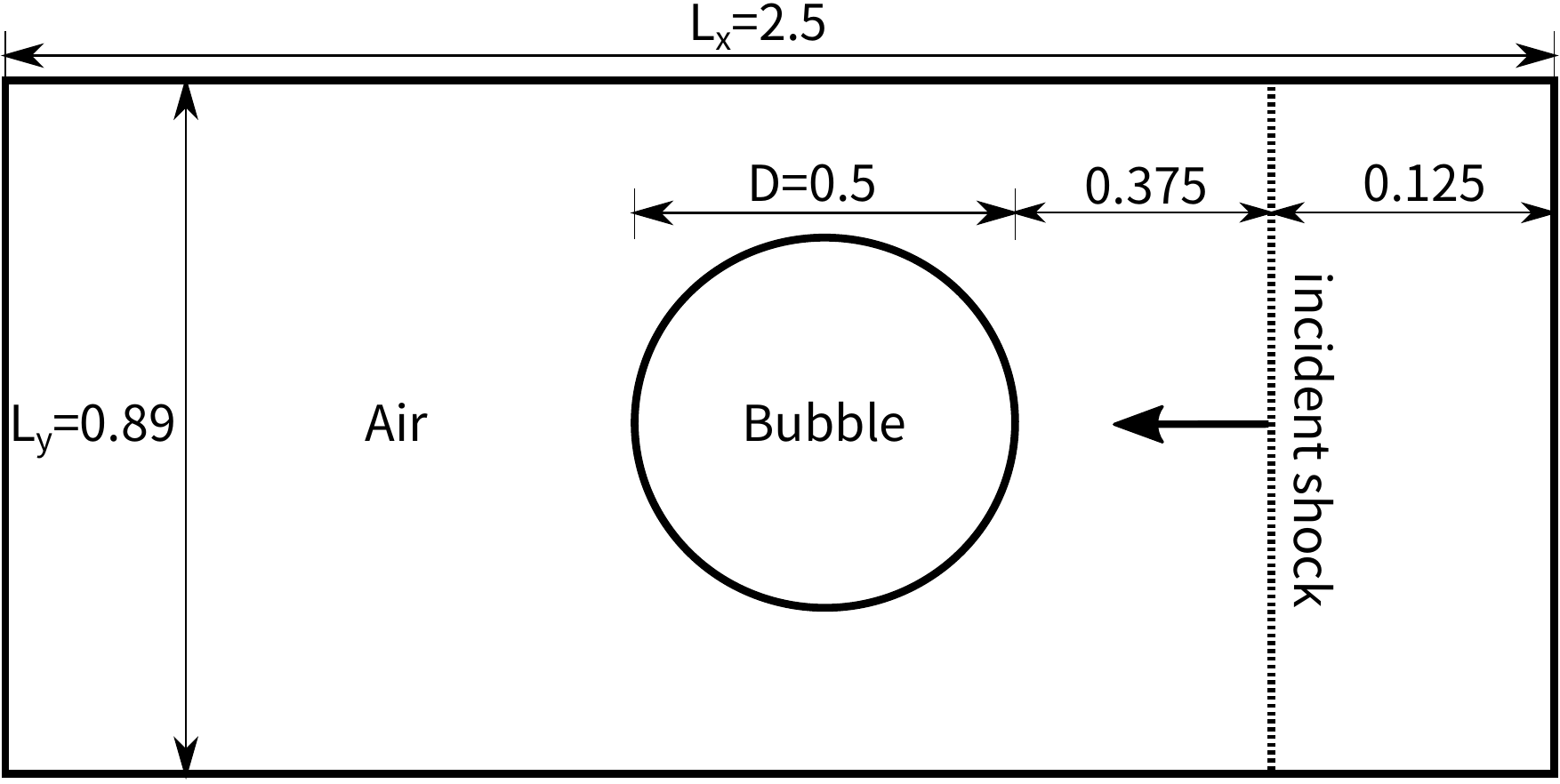}
\caption{\label{bubble}Diagram of the shock-bubble interaction problem}
\end{figure}

The fourth example is about the interaction problem of a planar shock wave with a cylindrical gas bubble. This problem is motivated by the experiments in Ref.~\onlinecite{haas_interaction_1987}, and some existing numerical simulations can be found in Refs.~\onlinecite{quirk_dynamics_1996,allaire_five-equation_2002,banks_high-resolution_2007,ton_improved_1996,fedkiw_non-oscillatory_1999}. In the experiments, a weak shock with the shock Mach number $M_s=1.22$ propagates from atmospheric air into a stationary cylindrical bubble filled with lighter helium or heavier Refrigerant 22(R22). The computational domain $[0,2.5]\times[0,0.89]$ composes of $2500\times 890$ square cells and the position of initial discontinuity is set in FIG.~\ref{bubble}.
The upper and lower boundaries are solid wall boundaries, whereas the left and right boundaries are non-reflective. The air outside and the gas inside the bubble are assumed initially to be in the temperature and pressure equilibrium, and the density and pressure of air outside the bubble are set to be unit.
\begin{table}[ht]
\caption{\label{parameters}Some parameters for the shock-bubble interaction problems in front of the shock wave}
\begin{ruledtabular}
\begin{tabular}{cddd}
Gas & \multicolumn{1}{c}{\mbox{Air}} &  \multicolumn{1}{c}{\mbox{Helium}+28\%\mbox{Air}} & \multicolumn{1}{c}{\mbox{R22}}\\
$\gamma$   &1.40   &1.648 &1.249\\
$C_v$   &0.72   &2.44 &0.365\\
$\rho$   &1   &0.182 &3.169\\
$p$ &1&1&1\\
$u$ &0&0&0\\
\end{tabular}
\end{ruledtabular}
\vspace{0.2cm} 
\end{table}
For the helium bubble case, the gas in the bubble is assumed as a helium-air mixture where the mass fraction of air is $28\%$, which is explained in Ref.~\onlinecite{haas_interaction_1987}. These materials are regarded as ideal gases, for which $\gamma$ and $C_v$ are taken from Ref.~\onlinecite{quirk_dynamics_1996} and presented in Table \ref{parameters}. Then the density of the gas inside the bubble is
\begin{equation*}
\frac{C_{v,air}(\gamma_{air}-1)}{C_{v,bubble}(\gamma_{bubble}-1)}.
\end{equation*}

\vspace{0.2cm}
 
\begin{figure}[htb]
\begin{minipage}[t]{0.45\textwidth}
(a)
\hfill
\includegraphics[width=0.9\textwidth]{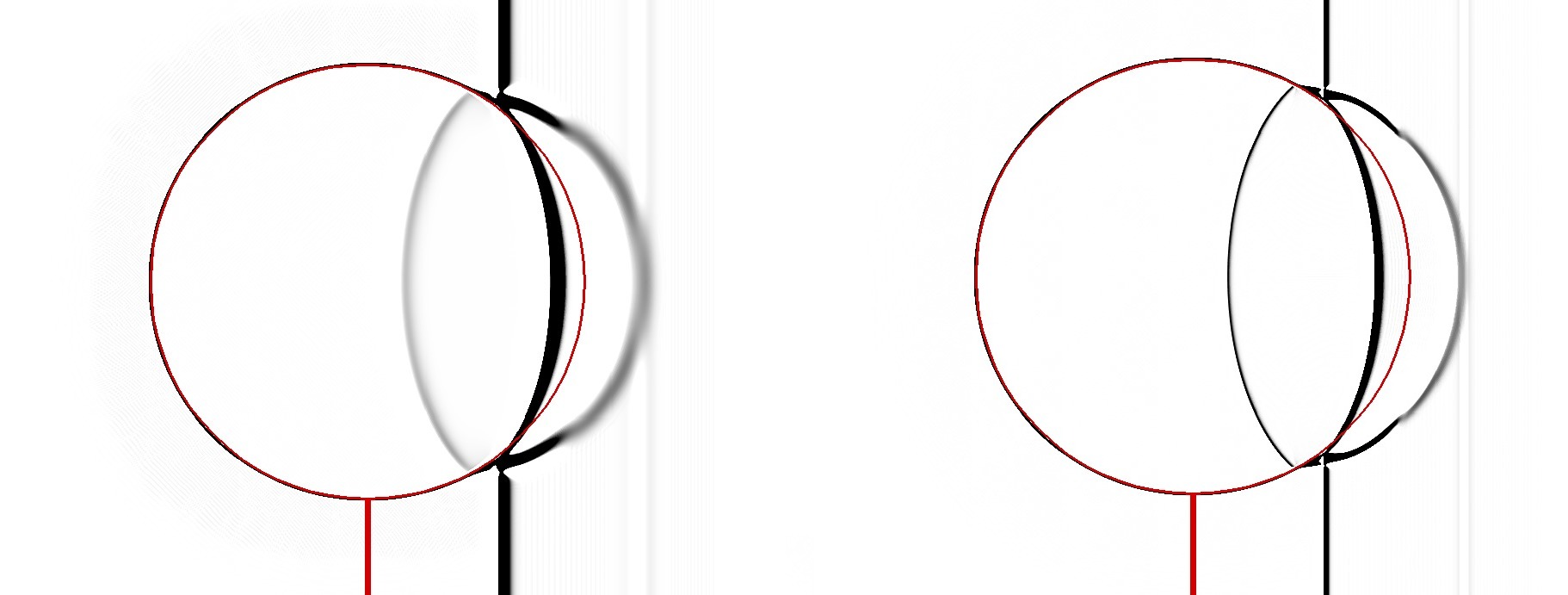}
\end{minipage}
\vfill
\begin{minipage}[t]{0.45\textwidth}
(b)
\hfill
\includegraphics[width=0.9\textwidth]{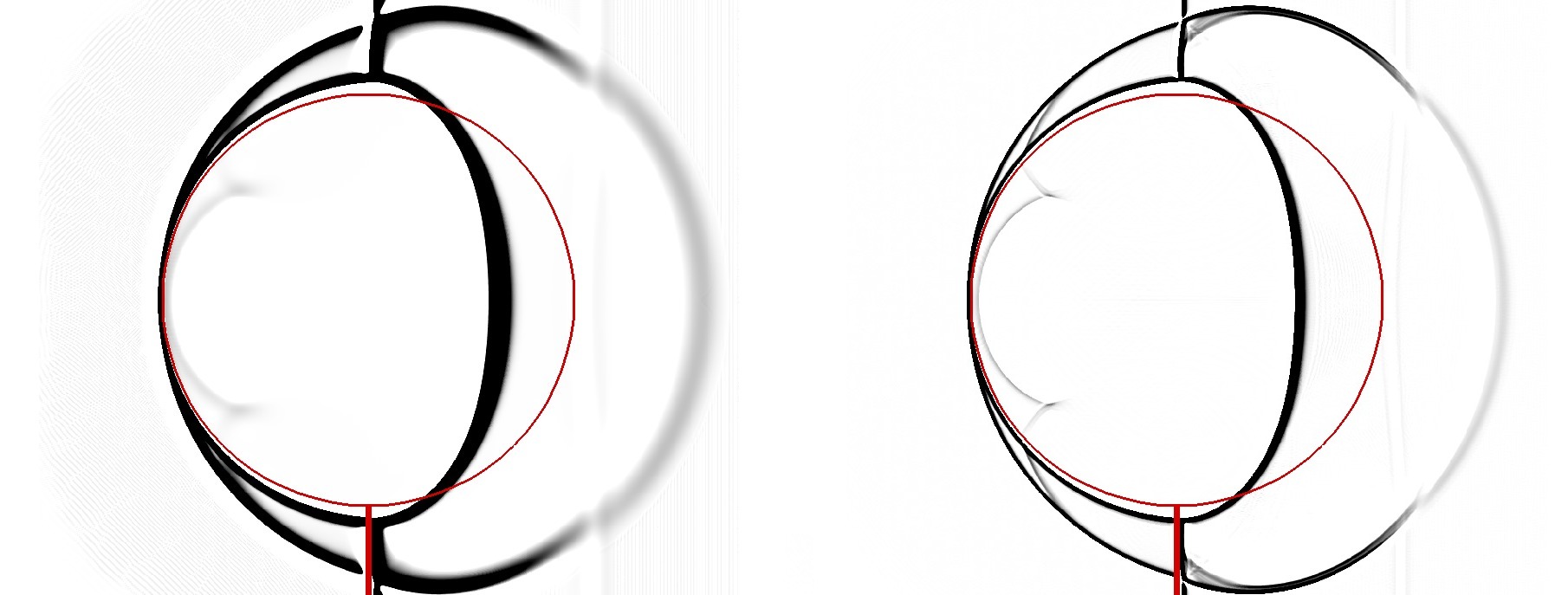}
\end{minipage}
\vfill
\begin{minipage}[t]{0.45\textwidth}
(c)
\hfill
\includegraphics[width=0.9\textwidth]{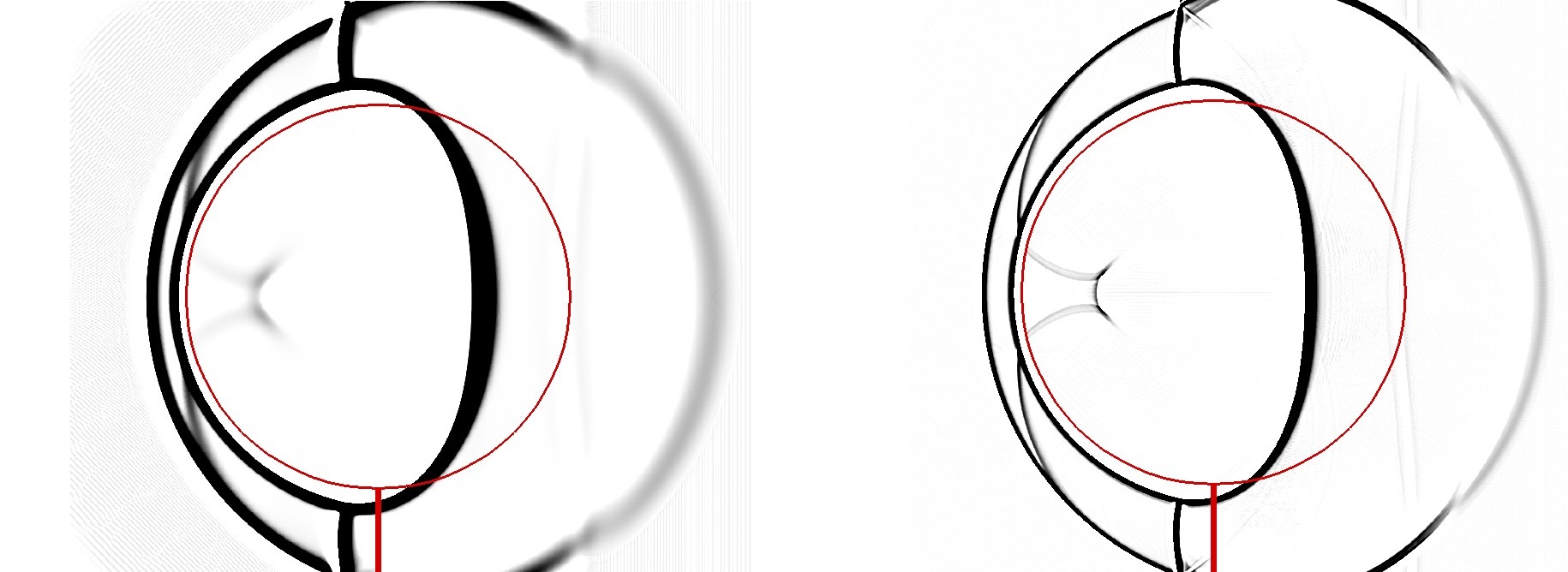}
\end{minipage}
\vfill
\begin{minipage}[t]{0.45\textwidth}
(d)
\hfill
\includegraphics[width=0.9\textwidth]{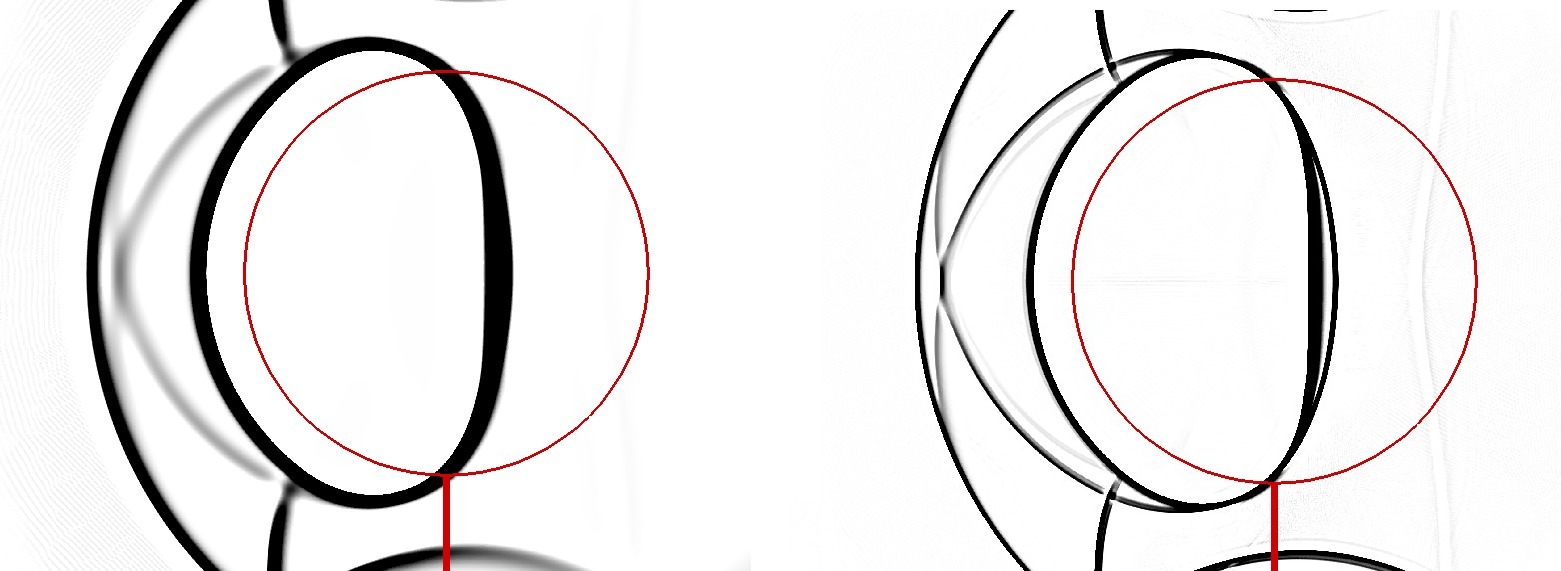}
\end{minipage}
\vfill
\begin{minipage}[t]{0.45\textwidth}
(e)
\hfill
\includegraphics[width=0.9\textwidth]{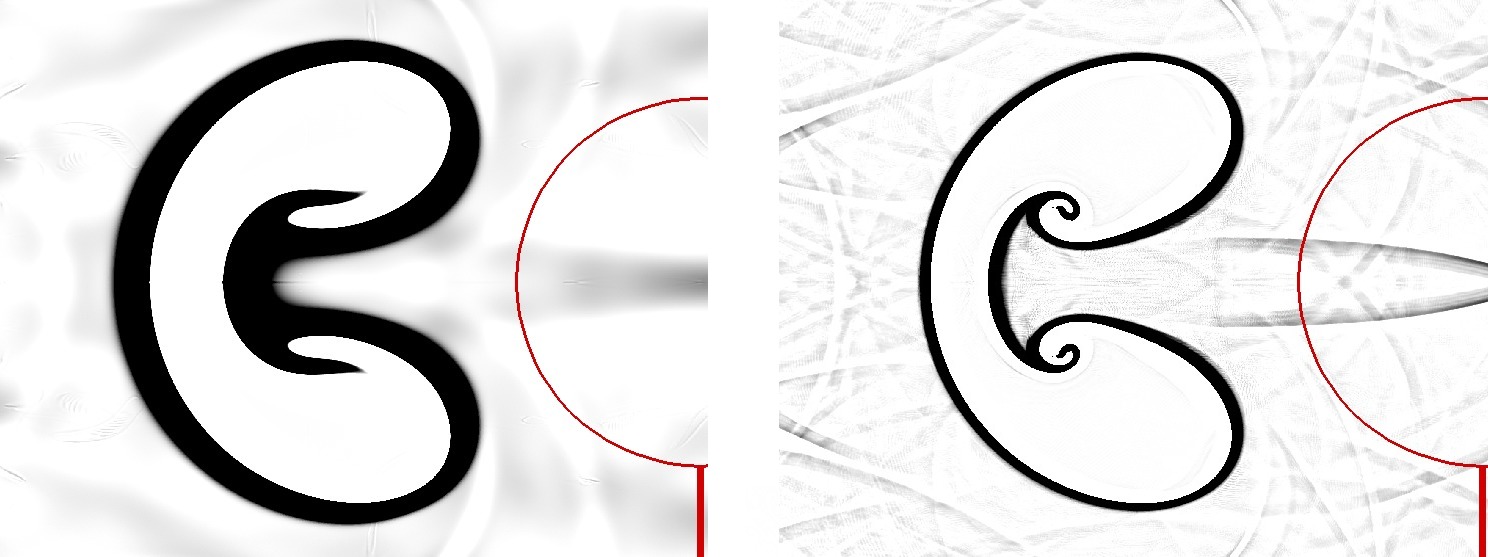}
\end{minipage}
\vfill
\begin{minipage}[t]{0.45\textwidth}
(f)
\hfill
\includegraphics[width=0.9\textwidth]{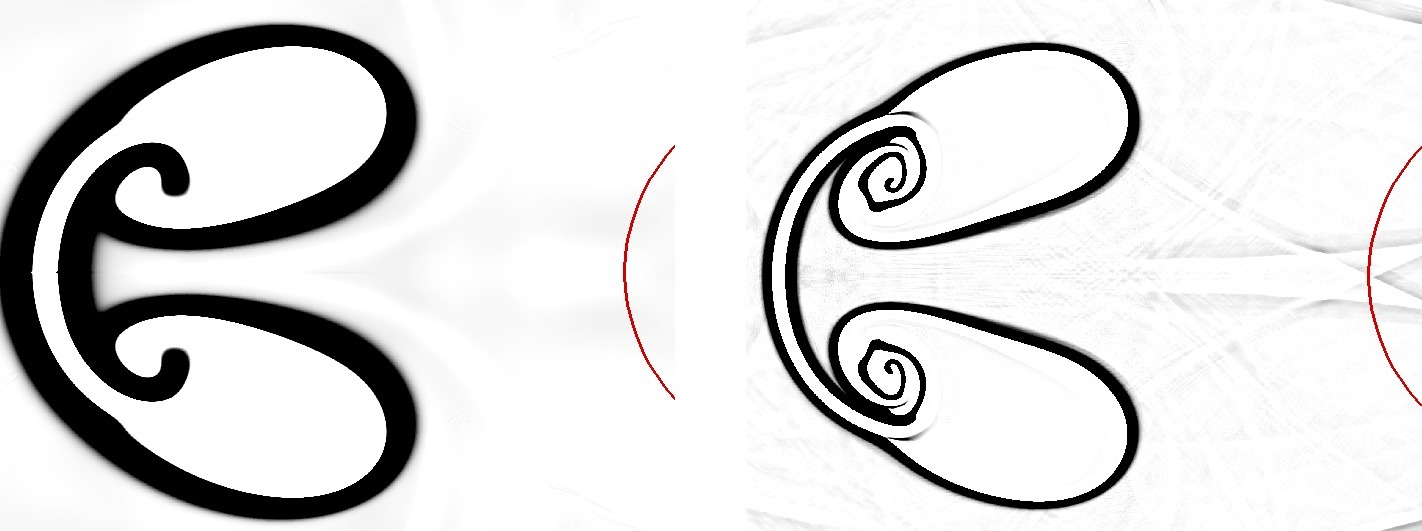}
\end{minipage}
\caption{\label{Helium}Numerical shadow-graph images of the shock-helium bubble interaction with $M_s=1.22$ obtained by ES-Godunov (left) and ES-GRP (right) at experimental times ($\mu$s): (a)$32$, (b)$62$, (c)$72$, (d)$102$, (e)$427$ and (f)$674$.  The intensity in shadow-graph images varies with the second derivative of density. The corresponding experimental  shadow-photographs can be found in Ref.~\onlinecite{haas_interaction_1987} (FIGURE 7).}
\end{figure}

\begin{figure}[htb]
\centering
\includegraphics[width=\linewidth]{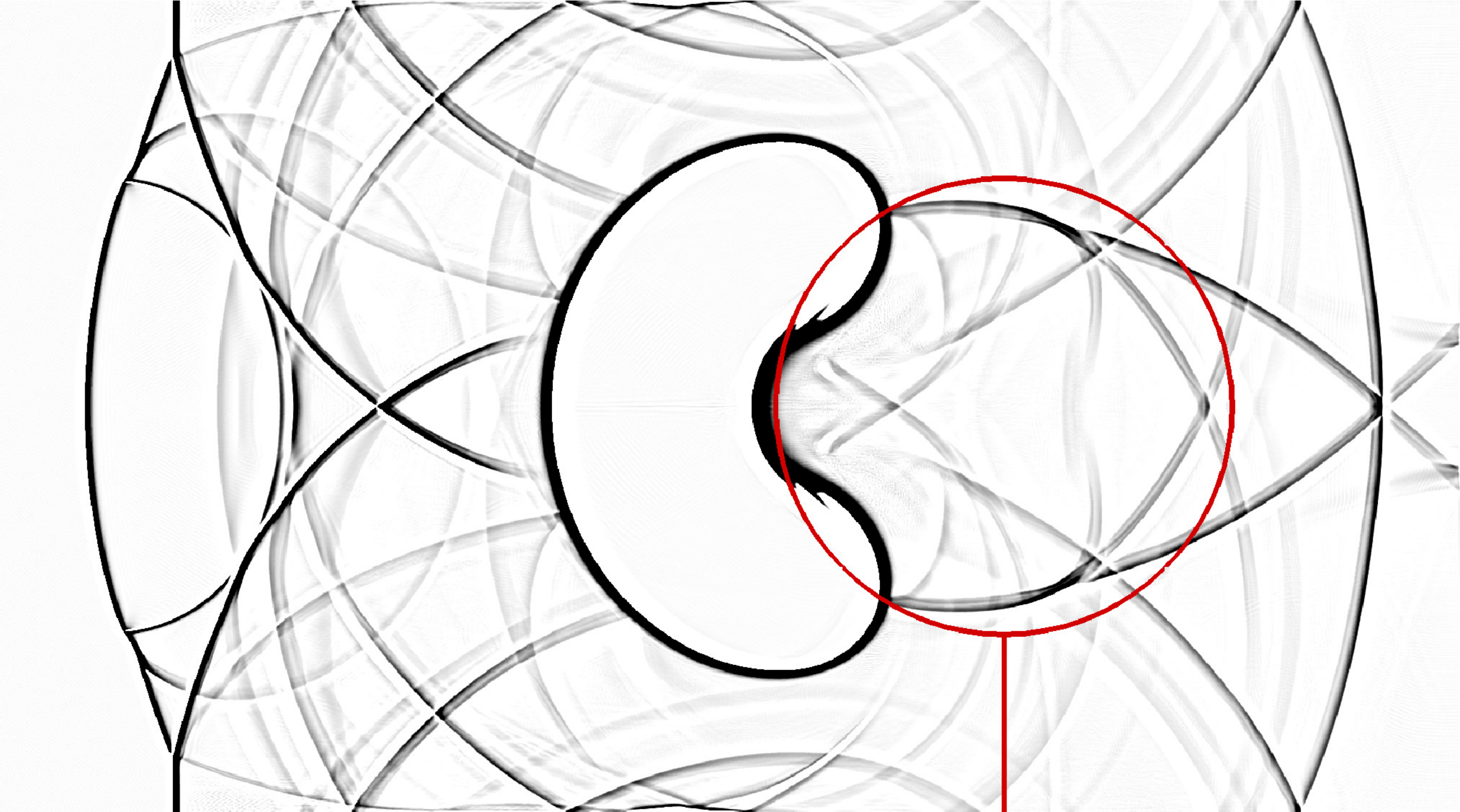}
\caption{\label{He_g}Numerical shadow-graph image of the shock-helium bubble interaction with $M_s=1.22$ by ES-GRP at experimental time $245\mu$s.}
\end{figure}

FIG.~\ref{Helium} compares the numerical shadow-graph images of the shock-helium bubble interaction problem by ES-Godunov and ES-GRP($\alpha=1$), corresponding to the experiments at different  times in Ref.~\onlinecite{haas_interaction_1987}. In order to better compare the results, the initial interface (red curves) is  added to the numerical shadow-graph images.  
FIG.~\ref{Helium}(a) shows the incident and reflected shock waves outside the bubble and a transmitted shock wave inside after the interaction between the shock and the right side of the bubble. Since the sound speed of the helium-air mixture inside the bubble is much greater than the sound speed of air outside, the transmitted shock wave propagates faster than the incident shock wave and reaches the left boundary of the bubble at experimental time $62\mu$s as shown in FIG.~\ref{Helium}(b). Then two secondary transmitted shock waves, connecting the primary transmitted shock wave and the left interface, are seen outside the bubble in FIG.~\ref{Helium}(c). The secondary transmitted shock waves intersect each other on the centerline and the internal reflected wave has diverged in FIG.~\ref{Helium}(d). Afterwards, the material interface continues to deform in FIGs.~\ref{Helium}(e) and (f). FIG.~\ref{He_g} shows the shadow-graph image of the whole flow field at $245\mu$s. It is observed that the primary transmitted wave is convex forward the helium bubble, which means the physical phenomenon that the helium bubble acts as a divergent lens for the incident shock.

\begin{figure}[htb]
\begin{minipage}[t]{0.45\textwidth}
(a)
\hfill
\includegraphics[width=0.9\textwidth]{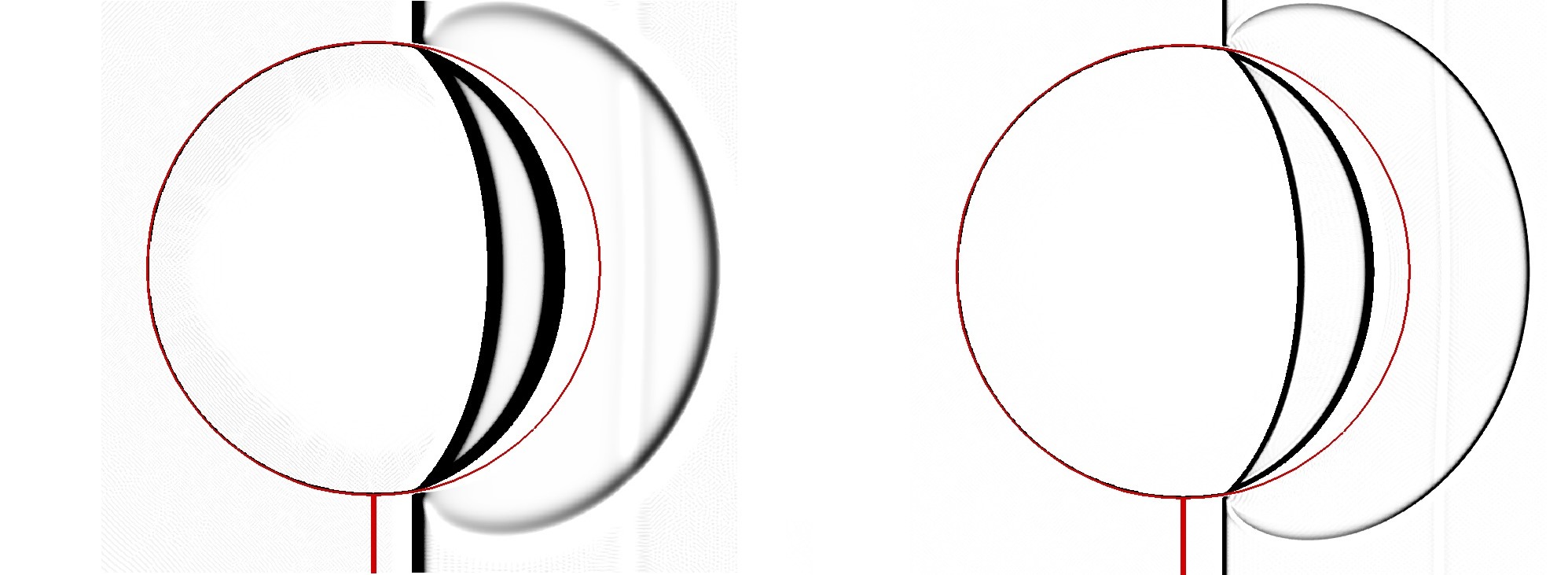}
\end{minipage}
\vfill
\begin{minipage}[t]{0.45\textwidth}
(b)
\hfill
\includegraphics[width=0.9\textwidth]{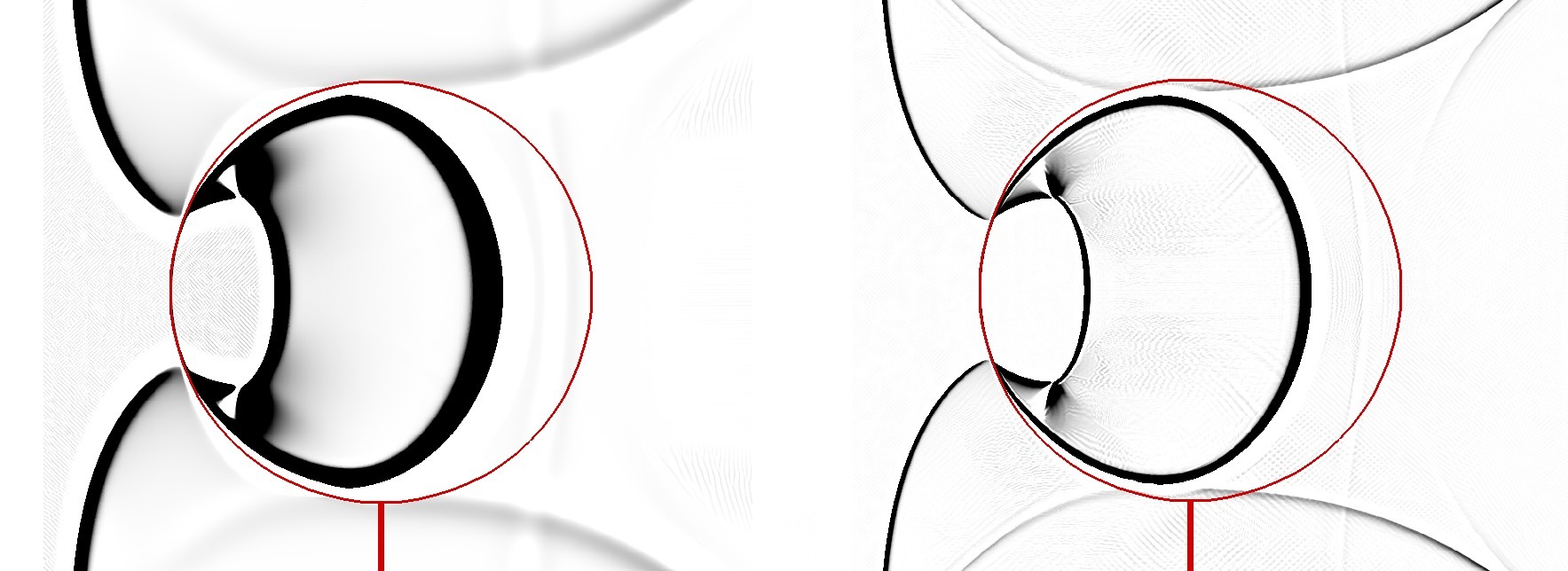}
\end{minipage}
\vfill
\begin{minipage}[t]{0.45\textwidth}
(c)
\hfill
\includegraphics[width=0.9\textwidth]{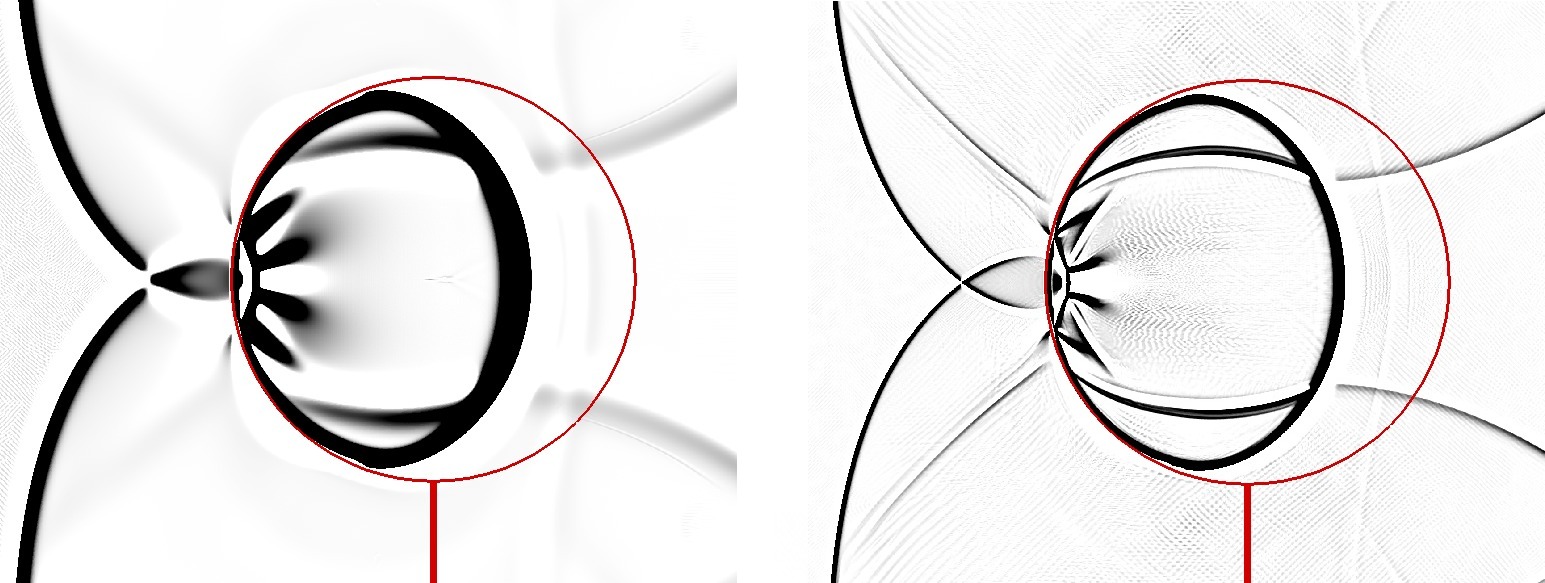}
\end{minipage}
\vfill
\begin{minipage}[t]{0.45\textwidth}
(d)
\hfill
\includegraphics[width=0.9\textwidth]{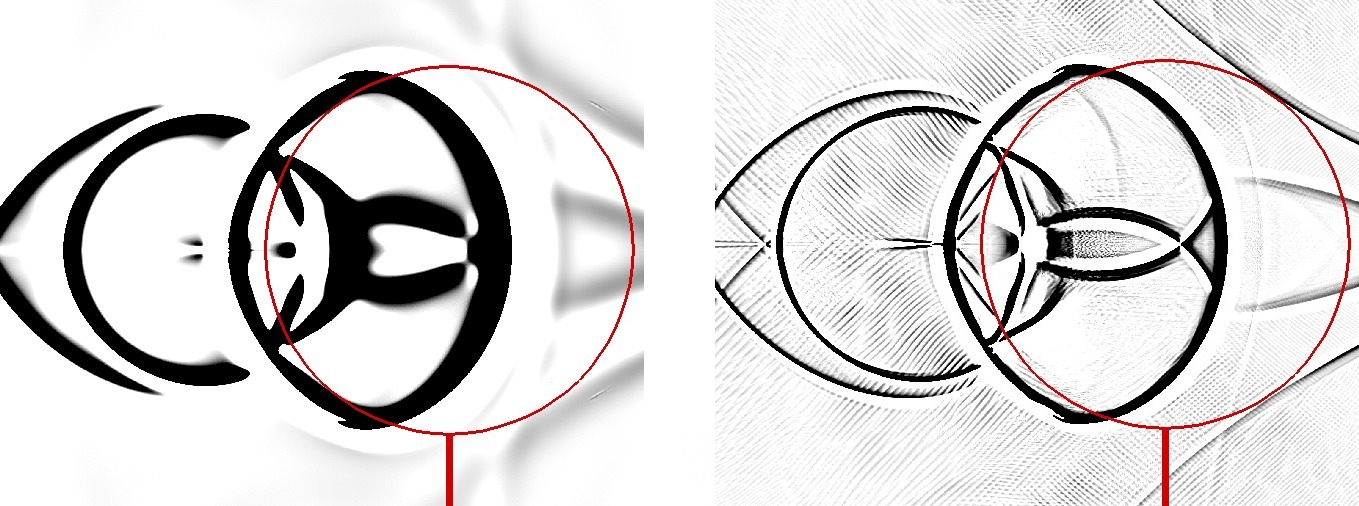}
\end{minipage}
\vfill
\begin{minipage}[t]{0.45\textwidth}
(e)
\hfill
\includegraphics[width=0.9\textwidth]{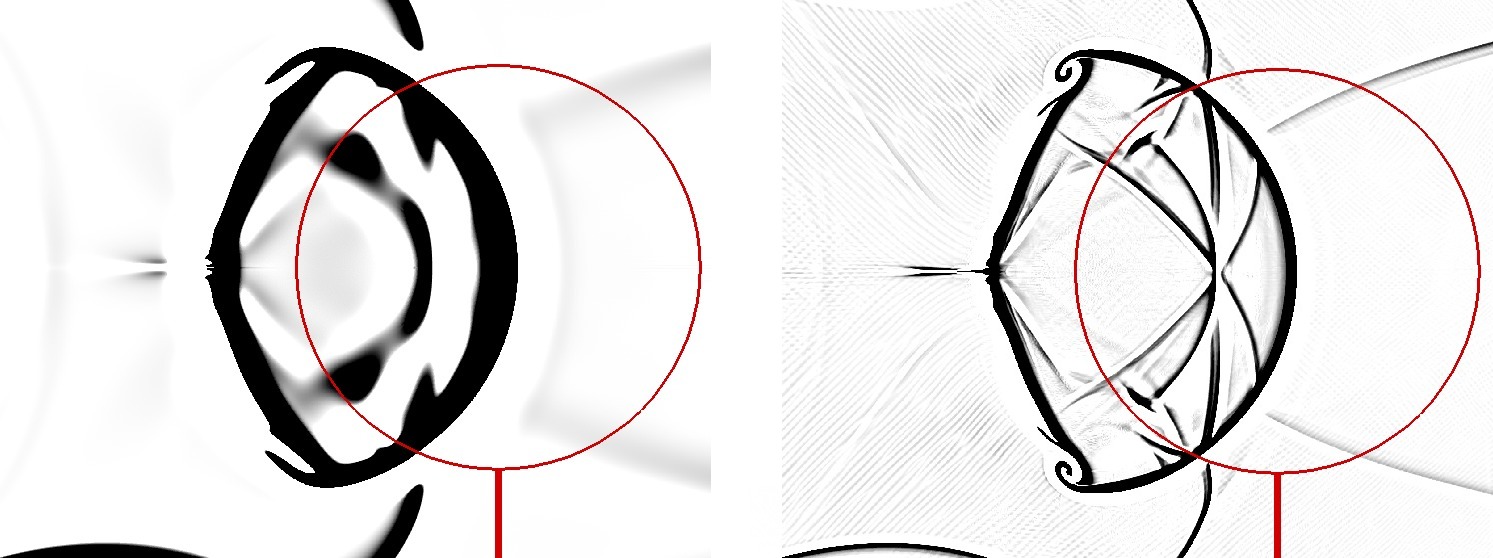}
\end{minipage}
\vfill
\begin{minipage}[t]{0.45\textwidth}
(f)
\hfill
\includegraphics[width=0.9\textwidth]{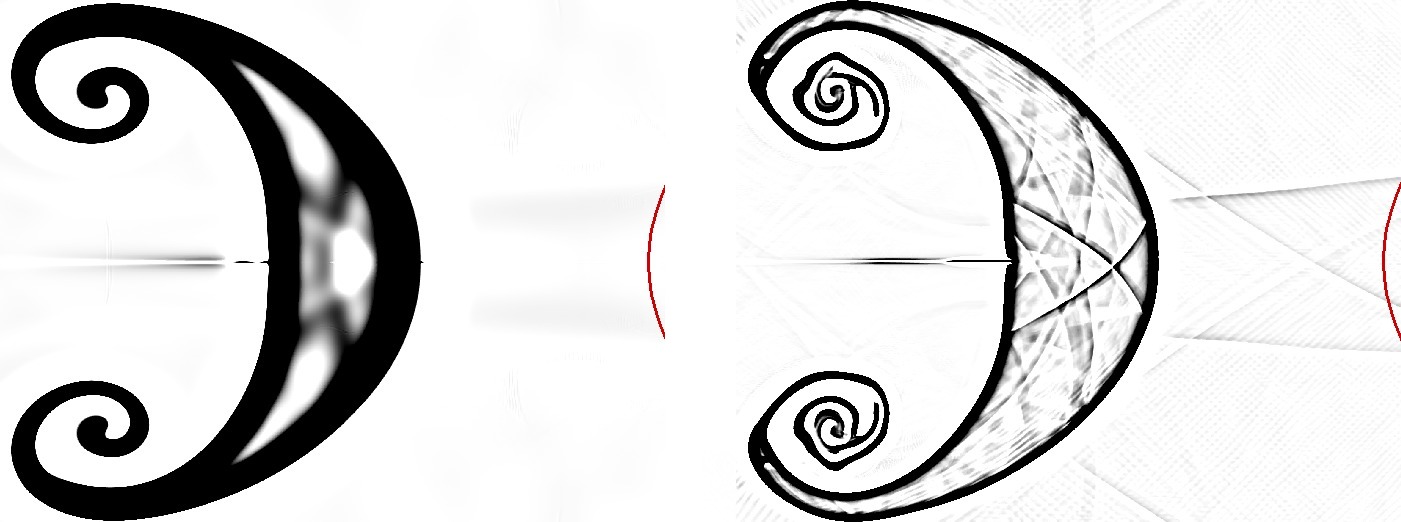}
\end{minipage}
\caption{\label{R22}Numerical shadow-graph images of the shock-R22 bubble interaction with $M_s=1.22$ obtained by ES-Godunov (left) and ES-GRP  (right) at experimental times ($\mu$s): (a)$55$, (b)$135$, (c)$187$, (d)$247$, (e)$342$ and (f)$1020$. The corresponding experimental  shadow-photographs can be found in Ref.~\onlinecite{haas_interaction_1987} (FIGURE 11).}
\end{figure}

\begin{figure}[htb]
\centering
\includegraphics[width=\linewidth]{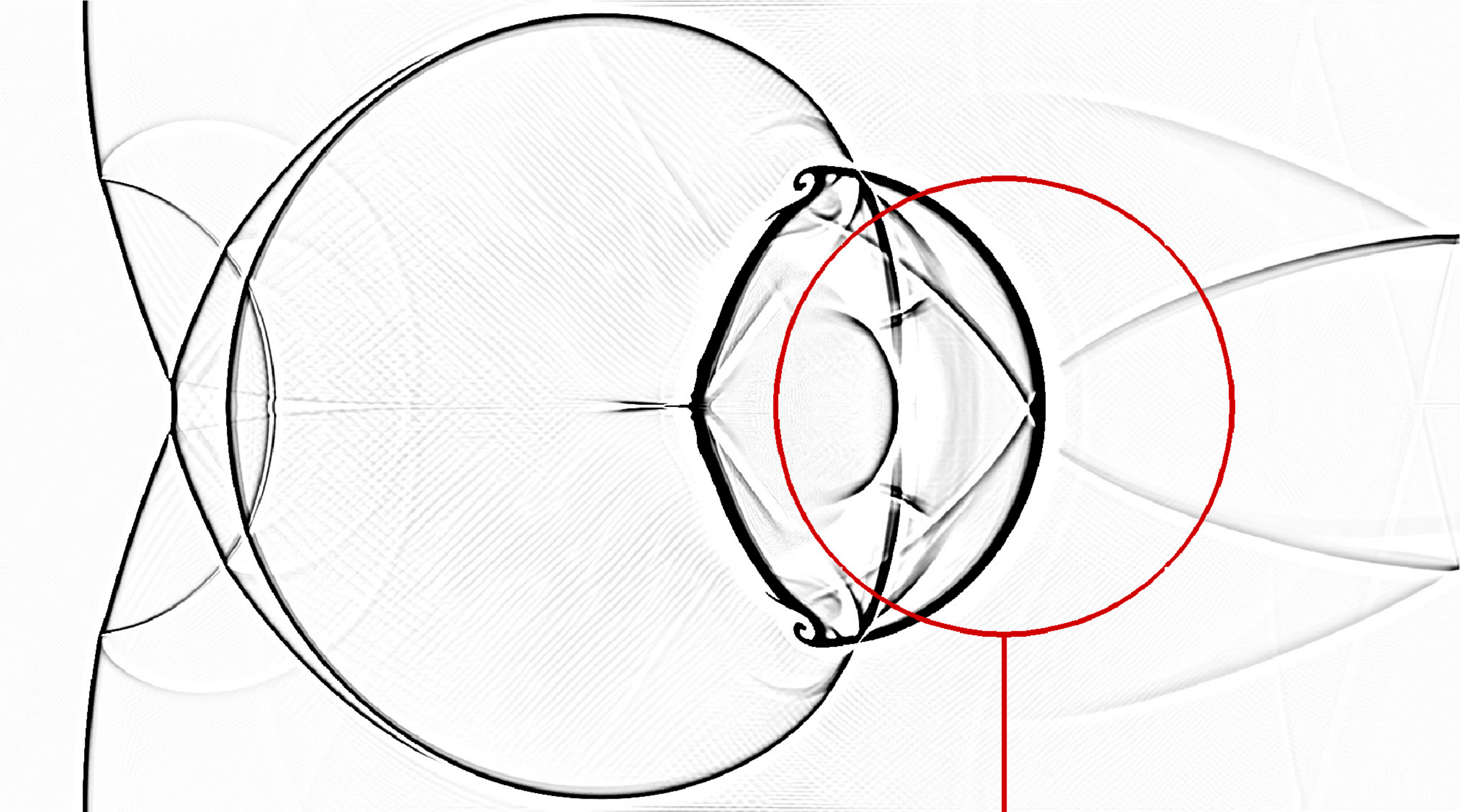}
\caption{\label{R22_f}Numerical shadow-graph image of the shock-R22 bubble interaction with $M_s=1.22$ by ES-GRP at the experimental time $318\mu$s.}
\end{figure}

\vspace{0.2cm}

Moreover, FIG.~\ref{R22} compares the numerical shadow-graph images of the shock-R22 bubble interaction problem by ES-Godunov and ES-GRP($\alpha=1$). FIG.~\ref{R22}(a) shows the incident and reflected shock waves outside the bubble and a refracted shock wave inside after  the interaction between the shock and the right side of the bubble. As the sound speed of R22 in the bubble is much smaller than the sound speed of air outside, the refracted shock wave inside the bubble propagates more slowly than the incident shock wave outside. In FIG.~\ref{R22}(b) the incident shock wave diffracts outside the cylinder and connects to the refracted wave inside the bubble.  After then, the two branches of the diffracted waves  cross each other  and the refracted shock focuses near the interface in FIG.~\ref{R22}(c). Then it expands radially outside the bubble in FIG.~\ref{R22}(d). High velocity created by the transmitted shock at its focus causes a central wedge to form on the downstream R22-air interface in FIG.~\ref{R22}(e). Finally, the interface deforms into a large vortex pair in FIG.~\ref{R22}(f). FIG.~\ref{R22_f} shows the shadow-graph image of the whole flow field at $318\mu$s. It is observed that the primary transmitted wave is concave forward the R22 bubble, which acts as a convergent lens for the incident shock. The numerical shadow-graph images show a very good agreement between the second-order numerical simulations and the laboratory experiments. As  ES-GRP is used, there are less instability along the material interface than the numerical results in Ref.~\onlinecite{quirk_dynamics_1996} and much clearer discontinuity surfaces are observed than those by  ES-Godunov. 

\begin{figure}[htb]
	$M_s=1.22$
	\begin{minipage}[t]{\linewidth}
		\includegraphics[width=\textwidth]{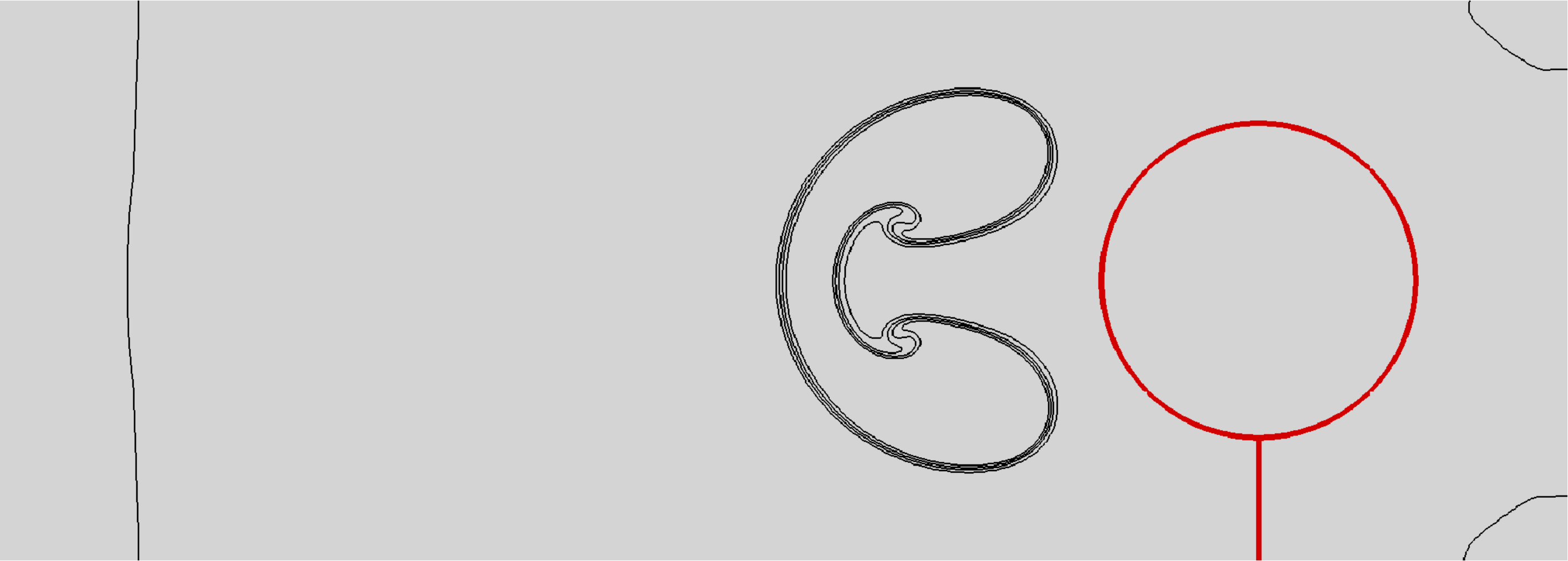}
	\end{minipage}
	\vfill
	$M_s=3$
	\begin{minipage}[t]{\linewidth}
		\includegraphics[width=\textwidth]{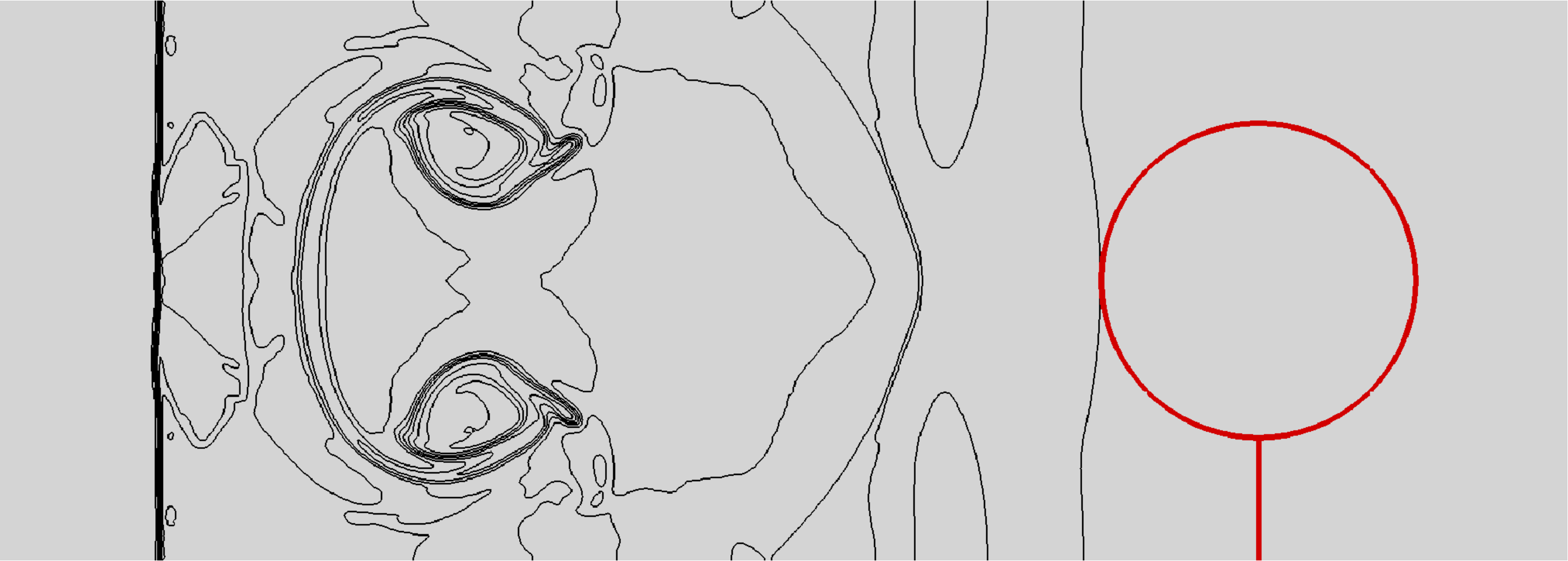}
	\end{minipage}
	\vfill
	$M_s=6$
	\begin{minipage}[t]{\linewidth}
		\includegraphics[width=\textwidth]{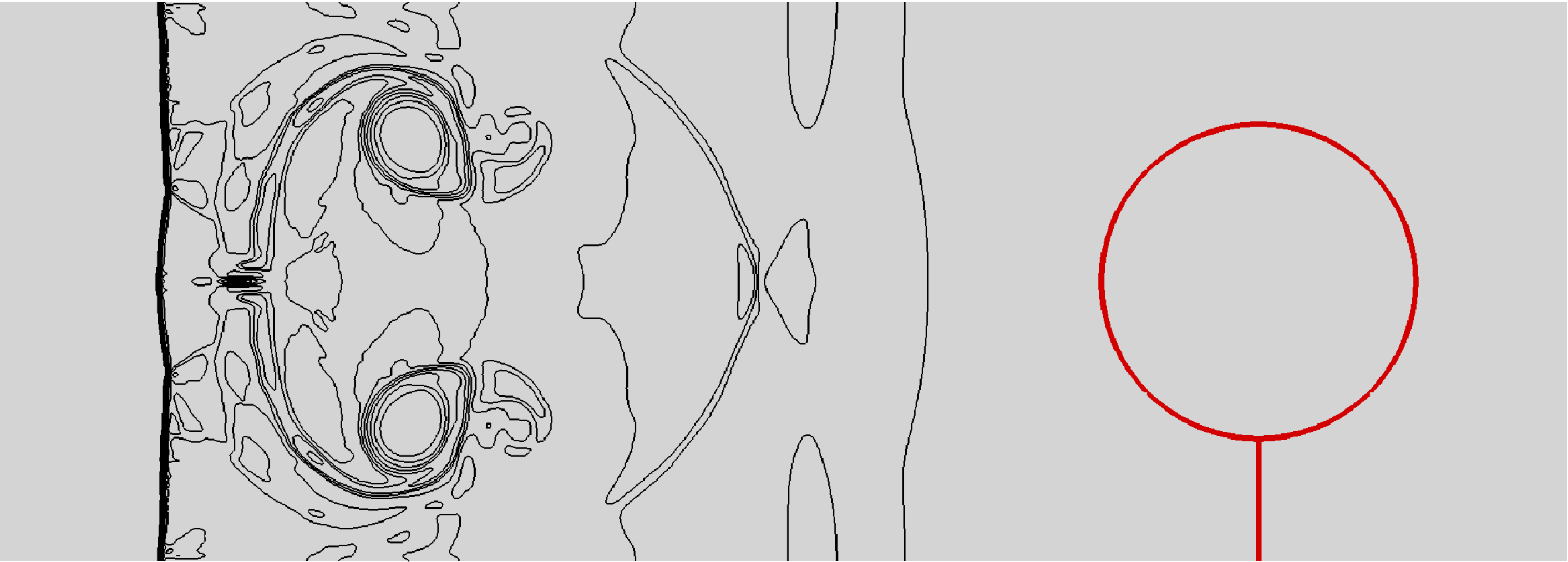}
	\end{minipage}
	\caption{\label{He_Mach}Density contours for the shock-helium bubble interaction by ES-GRP at $t/t_0=7.8$ (compare with Ref.~\onlinecite{bagabir_mach_2001}).}
\end{figure}

We continue to  simulate this shock-helium bubble interaction problem for various incident Mach numbers in the range of $1.22\leq M_s\leq 6$.  We use the same amount of grids ($560\times 200$) as the previous work in Ref.~\onlinecite{bagabir_mach_2001}. With the Mach numbers increasing, the accelerations of the bubbles also increase. Numerical results with $M_s=1.22,3,6$ by the ES-GRP($\alpha=1.5$) is presented in FIG.~\ref{He_Mach}. Using the definition of the time scale $t_0=R/(M_s\, c_{air})$, where $c_{air}$ is the sound-speed of ambient air and $R$ is the radius of the bubbles, we can compare the current numerical results with those in Ref.~\onlinecite{bagabir_mach_2001} at $t/t_0=7.8$, where the real computational time $t=0$ corresponds to the first impact of the shock with the bubble. We observe that differences are that there are clearer interfaces of the helium bubble with high $M_s$ by ES-GRP($\alpha=1.5$) and no oscillation is produced at the interface.

\subsection{Shock-accelerated gas cylinders}

\begin{figure}[htb]
\begin{minipage}[t]{0.49\textwidth}
\footnotesize{$~~~~~~~~t/t_0=0\quad t/t_0=28.8\quad t/t_0=48.9\quad t/t_0=75.7$}
\vfill
(a)
\hfill
\includegraphics[width=0.92\textwidth]{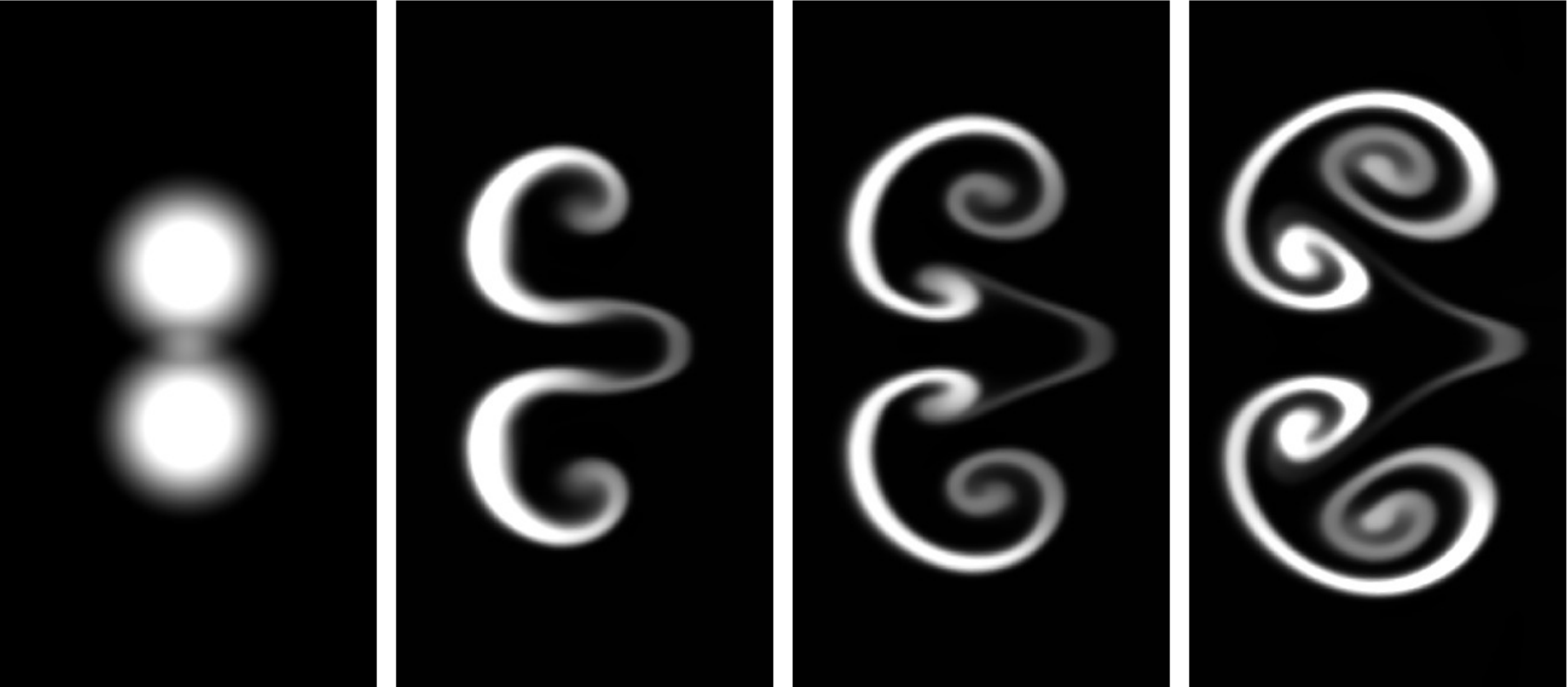}
\end{minipage}
\vfill

\begin{minipage}[t]{0.49\textwidth}
(b)
\hfill
\includegraphics[width=0.92\textwidth]{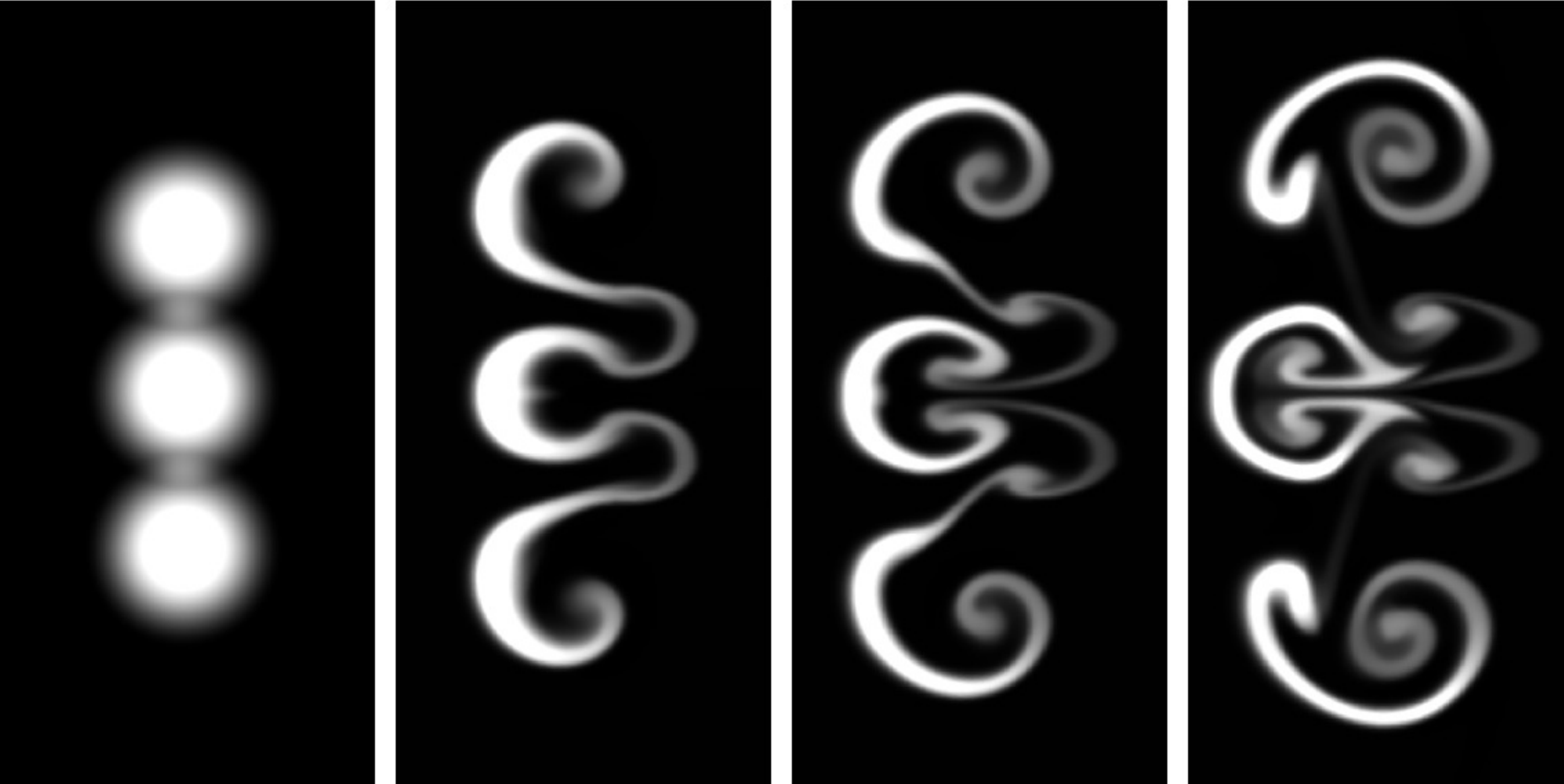}
\end{minipage}
\vfill

\begin{minipage}[t]{0.49\textwidth}
(c)
\hfill
\includegraphics[width=0.92\textwidth]{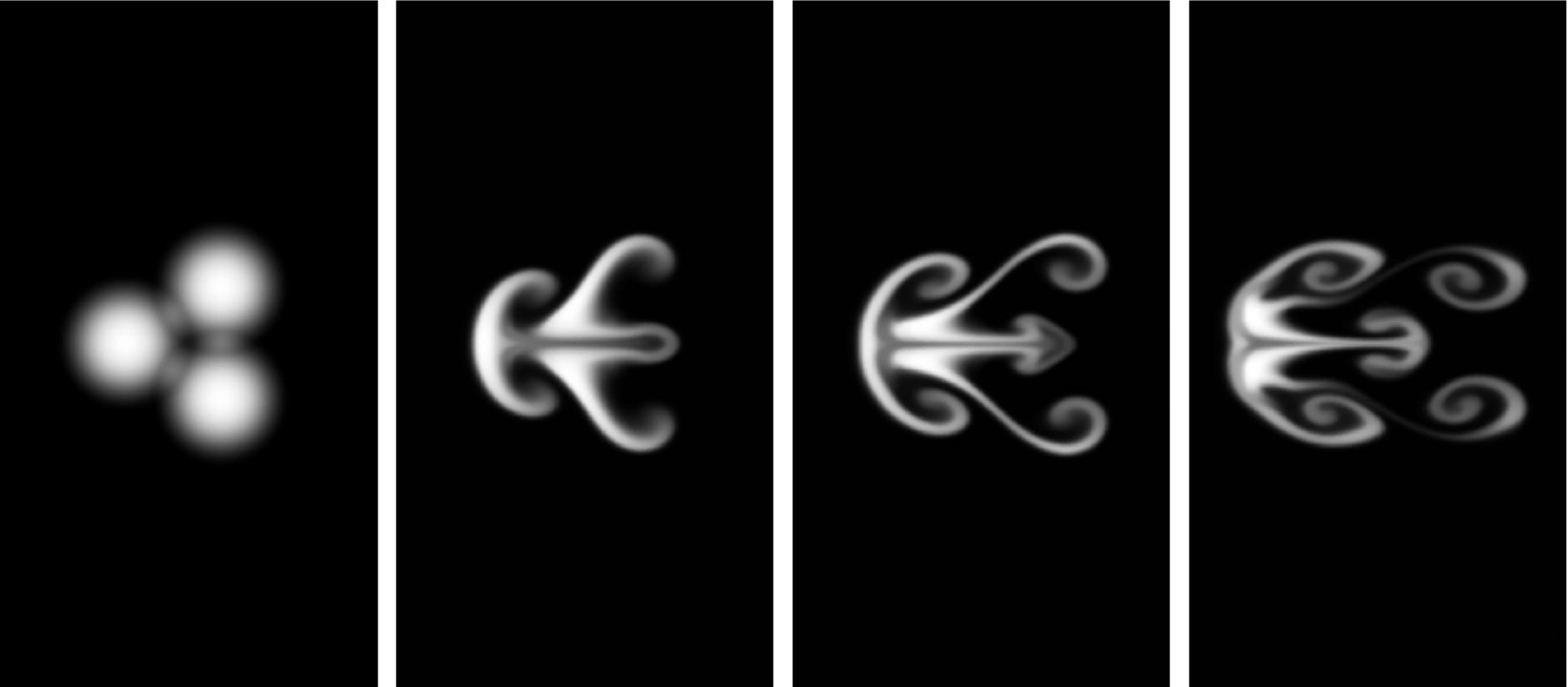}
\end{minipage}
\vfill
\begin{minipage}[t]{0.49\textwidth}
(d)
\hfill
\includegraphics[width=0.92\textwidth]{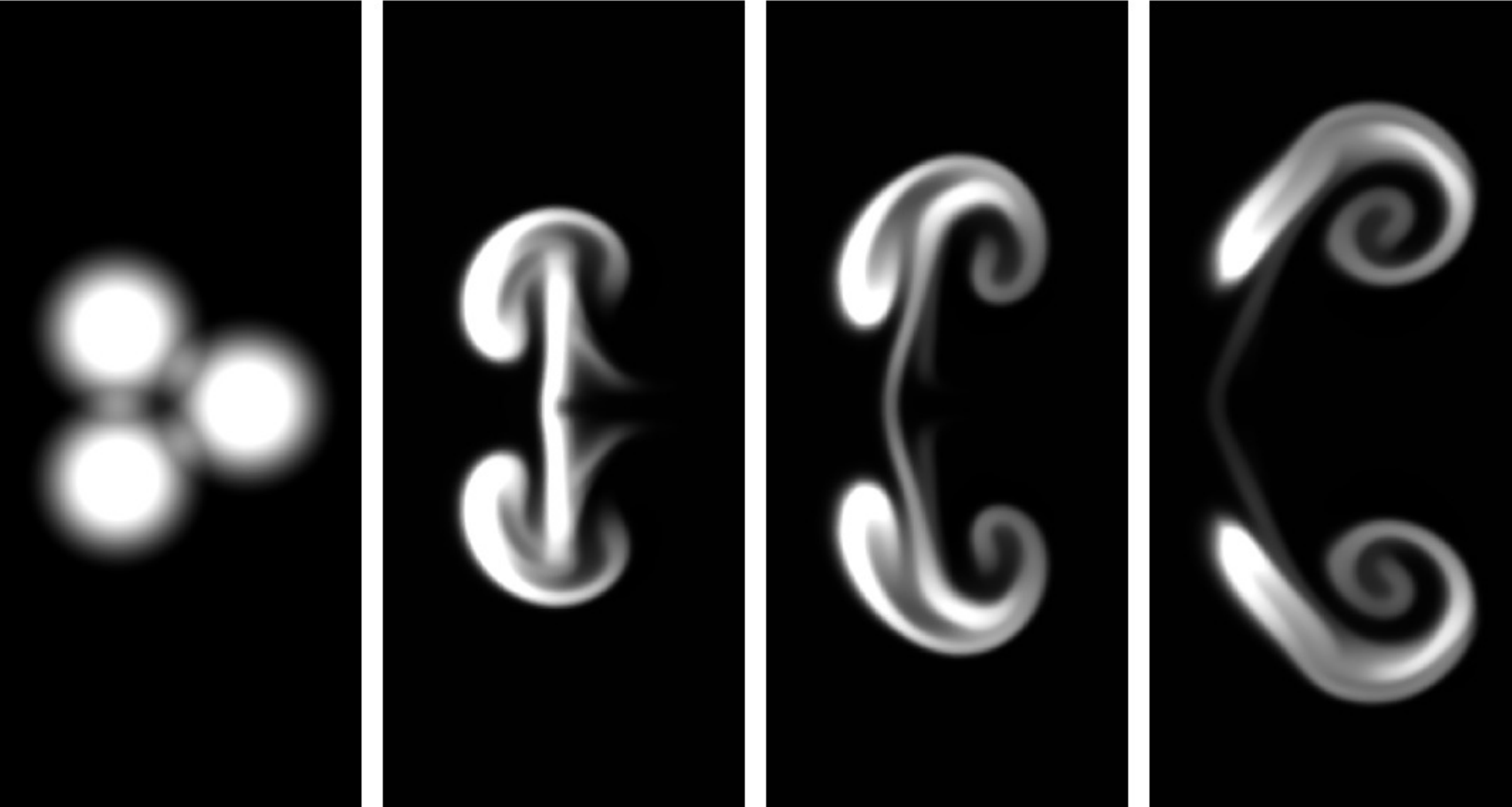}
\end{minipage}
\caption{\label{SF6}Gray scale images of density for the shock-accelerated $\mbox{SF}_6$ cylinders with $M_s=1.2$ by ES-GRP with different configurations at $t/t_0=0, 28.8, 48.9$ and $75.7$. The corresponding experimental images with planar laser-induced fluorescence (PLIF) can be found in Ref.~\onlinecite{kumar_stretching_2005} (FIGURES 8(a),9(a),10 and 11).}
\end{figure}

This non-oscillatory conservative scheme, effectively simulating the behavior of a single bubble, should be able to simulate multiple bubbles. Now we consider another example in Refs.~\onlinecite{tomkins_quantitative_2003,kumar_stretching_2005}, where a $M_s=1.2$ planar shock wave accelerates multiple gas SF$_6$ cylinders surrounded by ambient air.
The corresponding  numerical simulations for different shapes of initial configurations were displayed in Ref.~\onlinecite{kumar_complex_2007}.
The initial configuration of the gas cylinders located at $x=0.1$ is showed in FIG.~\ref{SF6} at the first instant of the shock and gas cylinders collision $t/t_0=0$, where $t_0$ is used to normalize the time. In the initial configuration of cylinders, the spacing between the centers of the cylinders is $S=1.5D$, where $D=0.031$ is the diameter of the cylinders at the experimental nozzle.
The numerical initial mass fraction of {SF}$_6$ in the circular mixing region is described by\cite{shankar_numerical_nodate}
\begin{align*}
&\phi_{SF_6}(r) = \\
&\left\{
\begin{aligned}
&\phi_{max}-\phi_{max}\exp\left[\frac{\left|\left(1-\frac{r}{R_d}\right)\pi\right|^{1.54}}{1.0082}\right], &~ |r|\leq R_d,\\
&0.0, &~ |r|>R_d,
\end{aligned}
\right.\nonumber
\end{align*}
fitting experimental measurement in Ref.~\onlinecite{tomkins_experimental_2008},
where $r$ is the distance to the center of the circle, $R_d=0.925D$, $\phi_{max}=0.83$ is the maximum mass fraction  of SF$_6$ measured before the shock impact.
We use the reverse computational domain in FIG.~\ref{bubble} with $[0,0.4]\times[0,0.4]$ composed of $400\times 400$ square cells and the position of initial shock is  located at $x=0.02$. The all boundaries are non-reflective. Air outside and gas mixture inside the circulars are assumed initially to be in atmospheric pressure, and the density and ratio of specific heats for air and SF$_6$ are set in Table \ref{parameters2}, which are taken from Ref.~\onlinecite{fan_numerical_2012}. To ensure that the SF$_6$ stays in the computational domain, a uniform velocity of pre-shock gases is set to $-0.43$.
\begin{table}[ht]
\caption{\label{parameters2}Some parameters for the shock-accelerated gas cylinders problems}
\begin{ruledtabular}
\begin{tabular}{cdd}
Gas & \multicolumn{1}{c}{\mbox{Air}} & \multicolumn{1}{c}{\mbox{SF}$_6$}\\
$\gamma$		&1.40		&1.094\\
$\rho$		&1.185		&5.971\\
$p$			&10.1325		&10.1325\\
$u$			&-0.43		&-0.43\\
\end{tabular}
\end{ruledtabular}
\vspace{0.2cm} 
\end{table}

In comparison with the experimental results in Ref.~\onlinecite{kumar_stretching_2005}, the numerical results by ES-GRP($\alpha=1.5$) in FIG.~\ref{SF6} shows perfectly consistent shapes of the density results, where the time scales $t_0$ are: (a)$6.34\times 10^{-3}$, (b)$5.55\times 10^{-3}$, (c)$4.48\times 10^{-3}$ and (d)$5.55\times 10^{-3}$.
In FIG.~\ref{SF6}(a), two vortex pairs are formed as a result of shock interaction with the two gaseous cylinders. At $t/t_0=48.9$, the material interface  starts to roll up inside the vortex cores.
Similarly in FIG.~\ref{SF6}(b), two vortices in opposite directions  form. The two  inner vortices are weaker than the outer vortices because the density gradients are smaller due to diffusion between the cylinders. 
In FIG.~\ref{SF6}(c), the inner vortices of the right two cylinders are weaker and the inner gas is pushed upstream.
In FIG.~\ref{SF6}(d), the gas in the left cylinder stretches in the span-wise direction, forming a bridge between the two outer vortex pairs. This bridge elongates with time, eventually breaks up.

\section{Conclusions}

The study of compressible multi-fluid flows is an important topic in theory, numerics and applications, which can be seen from the very incomplete references quoted here.  The researches were carried out in various ways such as physical experiments, physical modelings, numerical simulations and many others. In the present study, we focus on the analysis and design of numerical algorithms with numerical demonstrations based on a typical four-equation model. Certainly, the algorithms proposed here are compatible with the five-equation models \cite{allaire_five-equation_2002} and could be extended to more complex equations of state (EOS) \cite{menikoff_riemann_1989}.

The schemes we design are based on the Godunov scheme with an second order extension by using the GRP solver.  The positivity preserving of mass fractions and volume fractions is pivotal  as a numerical fluid mixing rule around  interfaces, for which  the hypotheses of equal partial pressures and no internal energy exchange are made and the exchange of kinetic energy is processed in the current scheme so that  no pressure oscillations arise from material interfaces, even though there is large density or temperature difference. Full conservation of our scheme can insure the correct simulation of shock waves or rarefaction waves near the material interfaces.   

A series of benchmark problems are tested  in order to  demonstrate the effectiveness and performance   of the current  method. The one-dimensional problems display  the better resolution of shock waves and the correct computation  of internal energy around material interfaces.  The two-dimensional shock-bubble interaction problems demonstrate  the performance of ES-GRP capturing  material interfaces, through the  comparison with the corresponding physical experiments. It is expected that this method can be applied to engineering problems practically. 

\begin{acknowledgments}
	Jiequan Li's research work is supported by NSFC with Nos.~11771054 and 11371063, and by Foundation of LCP.
\end{acknowledgments}

%

\end{document}